\documentclass[%
 reprint,
superscriptaddress,
nofootinbib,
 amsmath,amssymb,
 aps,
 pre,
longbibliography,
]{revtex4-1}
\usepackage[utf8]{inputenc}
\usepackage{siunitx}
\usepackage{upgreek}
\usepackage{amssymb}
\usepackage{amsmath}
\usepackage{booktabs}
\usepackage{float}
\usepackage{verbatim}
\usepackage{textcomp}
\usepackage{booktabs}
\usepackage{bbold}
\usepackage{xcolor}
\usepackage{graphicx}
\usepackage{epstopdf}
\usepackage{dcolumn}
\usepackage{bm}
\usepackage[colorlinks,
            linkcolor=red,
            anchorcolor=blue,
            citecolor=blue,
            urlcolor=blue
            ]{hyperref}

\begin{document}
\title{Driving quantum systems with repeated conditional measurements}

\author{Quancheng Liu }
\thanks{qcliu.ac@gmail.com}
\affiliation{Department of Physics, Institute of Nanotechnology and Advanced Materials, Bar-Ilan University, Ramat-Gan 52900, Israel}

\author{Klaus Ziegler}
\thanks{klaus.ziegler@physik.uni-augsburg.de}
\affiliation{Institut f\"ur Physik, Universit\"at Augsburg, $D-86135$ Augsburg, Germany}

\author{David A. Kessler}
\thanks{kessler@dave.ph.biu.ac.il}
\affiliation{Department of Physics, Bar-Ilan University, Ramat-Gan 52900, Israel}

\author{Eli Barkai}
\thanks{Eli.Barkai@biu.ac.il}
\affiliation{Department of Physics, Institute of Nanotechnology and Advanced Materials, Bar-Ilan University, Ramat-Gan 52900, Israel}

\date{\today}

\begin{abstract}

We investigate the effect of conditional null measurements on a quantum system and find a rich variety of behaviors. Specifically, quantum dynamics with a time independent $H$ in a finite dimensional Hilbert space are considered with repeated strong null measurements of a specified state. We discuss four  generic behaviors that emerge in these monitored systems. The first arises in systems without symmetry, along with their associated degeneracies in the energy spectrum, and hence in the absence of dark states as well. In this case, a unique final state can be found which is determined by the largest eigenvalue of the survival operator,  the non-unitary operator encoding both the unitary evolution between measurements and the measurement itself. For a three-level system, this is similar to the well known shelving effect. Secondly, for systems with built-in symmetry and correspondingly a degenerate energy spectrum, the null measurements dynamically select the degenerate energy levels, while the non-degenerate levels are effectively wiped out. Thirdly, in the absence of dark states, and for specific choices of parameters, two or more eigenvalues of the survival operator match in magnitude, and this leads to an oscillatory behavior controlled by the measurement rate and not solely by the energy levels. Finally, when the control parameters are tuned, such that the eigenvalues of the survival operator all coalesce to zero, one has exceptional points that corresponds to situations that violate the null measurement condition, making the conditional measurement process impossible.

\end{abstract}
\maketitle
\section{Introduction}\label{Introduction}

\begin{figure}
    \centering
    \includegraphics[width=1\linewidth]{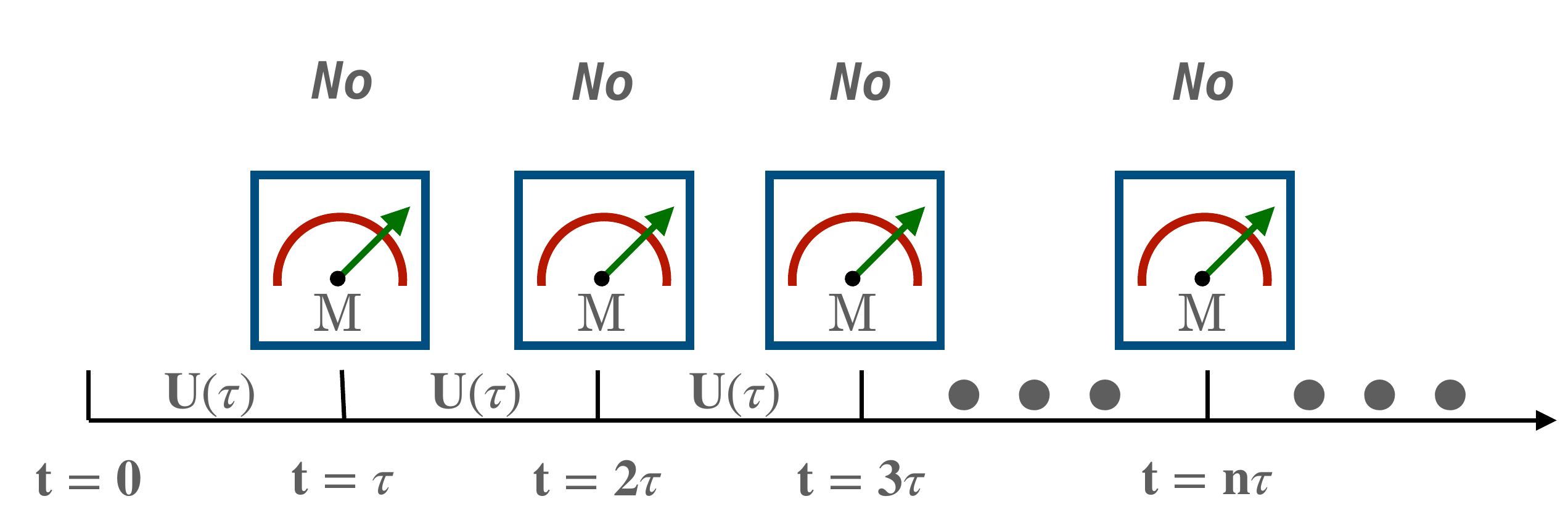}
    \caption{Schematic depiction of the quantum driving via null detections. The quantum system is initially prepared at some pure state. Local measurements of a target state $|\psi_{\rm d}\rangle$ are performed at discrete times $t= \tau, 2 \tau, \cdots, n \tau, \cdots$, whose outcomes are pinned down to null. We investigate the state of the quantum system under such repeated steering. }
    \label{fig:1}
\end{figure}

The concept of quantum steering \cite{RevModPhys.92.015001}, namely harnessing action at a distance to manipulate quantum states via measurements was introduced by Schr\"odinger \cite{cite-key,s1935,s1936}. Here we investigate theoretically the effect of repeated measurements on a quantum system. A finite dimensional Hilbert space is considered, and the Hamiltonian of the system is $H$. This can represent a tight-binding quantum walk on a graph, a two- or three- level atom, two entangled particles in the EPR setting \cite{RevModPhys.92.015001}, or a many body system. The system is prepared at time zero in a pure state $|\psi_{\rm in}\rangle $, and then every $\tau$ units of time we perform a {\em conditional} local measurement (see Fig. \ref{fig:1}). These conditional measurements drive the system, and the basic goal is to find  the behavior of the system in the long-time limit. It is shown below that one may find a rich panoply of dynamics that depend essentially on the sampling rate, the symmetry of $H$ and the presence or absence of degeneracy in the system. The possibility of implementing such ideas in the laboratory is made possible with the observation of single quantum trajectories \cite{PhysRevLett.56.2797,PhysRevLett.57.1696,PhysRevLett.57.1699,Plenio1998,Minev2019,Pokorny2020}, for example for superconducting qubits \cite{Murch2013}.

An example for a related  experiment is the quantum trajectory in  the process of shelving which can be used to reverse and reduce intermittency. Minev \textit{et al.} studied a superconducting  $V$-shaped artificial atom. Conditioning on {\em null} emission, the system is shelved in an energy level coupled to the ground state (see Fig. \ref{fig:three}, below). In the experiment, a feedback mechanism was used to prevent the excursions into the non-emissive state. Thus the conditional measurements, i.e. non-detection, in this case serve as a warning that the system is moving away from the bright state $|B \rangle$ defined in Fig. \ref{fig:three}. Similar experiments can be found in \cite{Murch2013,Weber2014,Tan2015,Foroozani2016}. These form remarkable demonstration of quantum trajectory concepts \cite{cite-key2,Dalibard1992,Gardiner1992,Carmichael1993,Mabuchi1996,Wiseman2002,Gambetta2008,Garrahan2018,Wiseman2009,Gammelmark2013,Guevara2015}. More recently the interplay between unitary dynamics and measurements have been emphasized \cite{Solfanelli2019,Buffoni2019,Mohammady2019,Riera-Campeny2020,Roy2020}, in the context of quantum computing. It was shown how the measurement rate can modify the emerging phases of the system \cite{Skinner2019,Nahum2020,Lavasani2021,Bao2021}. 

For a quantum system with a time independent Hamiltonian $H$ the energy spectrum determines the time evolution of the system, further energy is conserved. What will be the effect of a large number conditional null measurements? The latter are measurements where we do not detect the probed state. Not surprising in this process of measurement the energy is not conserved, and hence it is natural to wonder what it is in the long-time limit, similarly, in principle, for other observables. We will show that when $H$ has some symmetry built into it, the energy levels of the system do indeed control the long-time dynamics. However, not all the energy levels are contributing, instead only those energy levels that are degenerate. In this sense the mentioned three-level atom, which is non-degenerate can be considered as belonging to the next class of systems. In the absence of degeneracy, we find other physical scenarios. For example the system under conditions given below, can be driven to a unique state, like shelving  in the mentioned experiment. However, we identify this state in generality not as an energy state of the system, but rather it is an eigenstate of the survival operator, more specifically the state with the largest eigenvalue (see below). In other cases to be classified below, we find that the system exhibits periodicities. However, unlike unitary quantum mechanics, these oscillations are not determined by the energy levels alone. Instead the natural frequencies are controlled by the measurement rate as we shall show. Finally, there is a fourth class of systems, where we find exceptional points \cite{Kato1995,Heiss1999,Berry2004,Persson2000,Dembowski2001,Makris2008,Klaiman2008,Zhen2015,Cerjan2016,El-Ganainy2018}. The latter are well investigated, for example in the context of photonics  \cite{Goldzak2018,Bergholtz2019,Ozdemir2019,Bergman2021,Ben-Asher2020} and they describe in general degeneracies of non-Hermitian system \cite{moiseyev2011non,Ashida2020}. Here we show how at these points, corresponding to a special choice of sampling time $\tau$,  eigenvalues of the survival operator coalesce. Interestingly exceptional points are found for systems where the very basic condition of null measurements is not a possibility.  This means that when we have such exceptional points we cannot impose the condition in the first place. Instead we may work only close to this limit. 

At the heart of the analysis is the investigation of the left and right eigenvectors and eigenvalues of the survival operator.  This operator  describes the unitary evolution followed by the local measurement.  This operator has been previously investigated in the context of the quantum first detection problem \cite{Krovi2006,PhysRevA.74.042334,Varbanov2008,Gruenbaum2013,Grunbaum2014,Dhar2015,FriedmanE,Thiel2018,Thieldark}. Roughly speaking, in the latter case one is interested in the statistics of the first detection event, namely when do we detect the system for the first time? Here we are interested in the state of the system itself after many measurements conditioned that we never detect. Similar to the original idea of Schr\"{o}dinger, we consider two sets of measurements, the first is a set of many local measurements, this then drives the system to some state. The second measurement is of another operator, say $H$ itself, and it is neither conditioned nor local. Now if we had $n$ null measurements, prior to the final one, then obviously this influences the statistics of the final unconditional measurement.  The focus on the final state of the system is motivated in part from our interest in statistical physics, where the long-time limit of a driven system is of special interest, and energy is the natural choice of observable. 

 The manuscript is organized as follows. Secs. \ref{sec:dark} and \ref{sec:charge bio}  give the model, general discussion of the mathematical properties of the survival operator, a mapping of the quantum problem to classical charge theory, and a simple discussion on the exceptional points. In Secs. \ref{sec:formal} and \ref{sec:fix and dynamics} we present the main results, namely the state function under conditional measurements, for finite and infinite measurement times respectively.  In Secs. \ref{Sec:three} and \ref{sec:tree}, we present three examples, a three-level system, an artificial atom that can shelve, and a glued binary tree with degeneracy. In Sec. \ref{Sec:charge}, we present general insights on the final state of the system with a perturbation approach. Then in Sec. \ref{sec:ep}, we discuss exceptional points in a more general setting from the point of view of the symmetry of charge configuration.   We close the manuscript with a summary in Sec. \ref{sec:summary}.

\section{The survival operator}\label{sec:dark}

\begin{figure}
    \centering
    \includegraphics[width=0.7\linewidth]{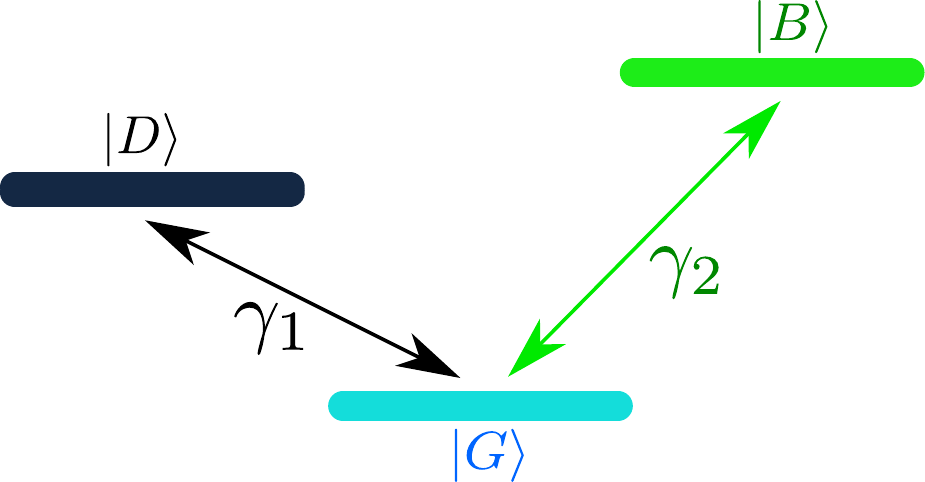}
    \caption{Three-level $V$-shaped system. We initially prepare the system in the ground state $|G\rangle$.  The matrix element of $H$  describing  jumping between $|G \rangle$ and $|D\rangle$ ($|B\rangle$) is $\gamma_1$ ($\gamma_2$), where $\gamma_2 \gg \gamma_1$. The detection state is the state $|B\rangle$.}
    \label{fig:three}
\end{figure}

As mentioned, we are interested in the evolution of the wave function of a system conditioned on null measurements. The model describes unitary dynamics interleaved with measurements. The system is discrete and finite, and $H$ is the Hamiltonian, which governs the dynamics between measurement events.  At the times $\tau$, $2\tau$, $3\tau$, $\cdots$ we perform measurements in an attempt to detect the system in the target state $|\psi_{\rm d}\rangle$.  For example, for a tight-binding quantum walk on a tree (see Fig. \ref{fig:G4tree}), discussed below,  $|\psi_{\rm d}\rangle$ describes a localized node on the graph, or $|\psi_{\rm d}\rangle$ can be an energy level (see Fig. \ref{fig:three}), etc. $\tau$ is a free parameter. The outcome of a given measurement, at least in principle, is either the system is detected at the target state $|\psi_{\rm d}\rangle$ or not. We condition the measurements and consider only the realization of the process described by the string: no, no, $\cdots$, no, which repeats $n$ times, namely the particle is never found in the target state, e.g. the quantum walker is not detected on the specified node of the graph.

 Therefore, the wave function after the first measurement is $|\psi_1\rangle = N_1(\mathbb{1} -D)\hat{U}(\tau)|\psi_{\rm in}\rangle$. Here, $D= |\psi_{\rm d}\rangle\langle \psi_{\rm d}|$ is the projection operator onto the measured state, while $\hat{U}(\tau) = \exp(-iH \tau)$ is the unitary evolution operator, and we have set $\hbar=1$. The operator
\begin{equation}
	\hat{\mathfrak{S}} =(\mathbb{1} -D)\hat{U}(\tau)
	\label{eq:1}
\end{equation}
incorporates the unitary evolution in the time interval $\tau$ followed by the projection described by $\mathbb{1}-D$. The latter, when acting on the state function, wipes out the amplitude on the detected state, as the result of the measurement is null. The operator $\hat{\mathfrak{S}}$ is called the survival operator as it describes the wave function that survives the detection (if the particle had been detected, the wave function would have collapsed on $|\psi_{\rm d}\rangle$). Clearly, it is a non-unitary operator. Immediately after the $n$th measurement, the wave function $|\psi_n \rangle$ reads:
\begin{equation}
	|\psi_n \rangle= N_n \hat{\mathfrak{S}}^n |\psi_{\rm in}\rangle.
	\label{eq:waveG}
\end{equation}
Below we suppress the index $n$ in the normalization $N_n$.

We are interested in the large-$n$ limit. In this limit, we encounter a variety of dynamical features explored below. In one case, the dynamics are determined by a sub-set of energy levels, in another by a fixed point where the system has no dynamics at all. Finally,  we observe oscillatory behavior; however, unlike the first case scenario, the eigenvalues determining this behavior are not standard energy levels of $H$. They are controlled by $H$, $D$, and $\tau$. Our goal is to classify these behaviors and see when they emerge. What is clear is that the large-$n$ limit is described by the set of eigenvalues of $\hat{\mathfrak{S}}$ of largest magnitude; hence this is what we study next.

As usual, since the operator $\hat{\mathfrak{S}}$ is non-Hermitian it has left and right eigenvalues and eigenvectors and these are denoted
\begin{equation}
	\hat{\mathfrak{S}}|\xi^R\rangle = \xi |\xi^R\rangle, \quad \langle \xi^L|\hat{\mathfrak{S}} =\langle \xi^L|\xi.
	\label{eq:eigenG}
\end{equation}
Here we used the fact, well known from linear algebra, that the left and right eigenvalues are equal; hence they are denoted with $\xi$. These are complex numbers, and for the problem at hand, it was shown that these eigenvalues are all on the unit disk $|\xi|\leq 1$ \cite{PhysRevA.101.032309,Thieldark}. We will use  normalized states $\langle \xi^R|\xi^R \rangle= \langle \xi^L|\xi^L\rangle=1$.

We start the analysis by considering three types of eigenvalues. The case $\xi=0$, the case $0 \textless |\xi| \textless 1$, and then $|\xi|=1$. When the eigenvalue is zero, we have
\begin{equation}
	|\xi^R\rangle = \hat{U}^{-1}|\psi_{\rm d}\rangle, \quad \langle \xi^L|=\langle \psi_{\rm d}|,
\end{equation}
which is obvious due to $(\mathbb{1}-D)U U^{-1}|\psi_{\rm d}\rangle =0$ and $\langle \psi_{\rm d}|(\mathbb{1}-D)=0$. The physical meaning of the $|\xi^R\rangle$ state is clear, as it is a state that is detected with probability one in the first measurement attempt. In fact, if we start in this state, namely $|\psi_{\rm in}\rangle = U^{-1}|\psi_{\rm d}\rangle$ we cannot achieve the desired conditional measurement, since the first measurement is always a successful. This implies that not all initial conditions can yield a string of $n$  null measurements, an issue we will return to below. 

More profound are the states with eigenvalues $|\xi|=1$ on the unit circle. This set can be empty as we show below, but for now we assume that such states exist. These right eigenstates satisfy
\begin{equation}
	\hat{\mathfrak{S}}|\xi^R\rangle= \underbrace{\exp(i \theta) }_{\xi}|\xi_R\rangle.
\end{equation}
We express these states in terms of linear combinations of stationary states of the time independent Hamiltonian $H$. For that denote $H|E_{k,l}\rangle=E_k|E_{k,l}\rangle$, where $k$ is the index of the distinct energy levels, while $l$ is a quantum number denoting the degenerate sub-levels, so that $l=1, \cdots, g_k$, where $g_k$ is the degeneracy of the energy level $E_k$. Degeneracy and hence symmetry play a crucial role here.

An obvious eigenstate of $\hat{\mathfrak{S}}$  is an energy state $|E_{k,l}\rangle$ which is orthogonal with respect to the detected state, namely if we have in our system a state $\langle \psi_{\rm d}|E_{k,l}\rangle=0$ we get
\begin{equation}
	(\mathbb{1}-D)\hat{U}(\tau)|E_{k,l}\rangle =\exp(-i E_k \tau)|E_{k,l}\rangle.
\end{equation}
Hence all energy states which are orthogonal to the detected state are right eigenvectors of the survival operator $|\xi^R\rangle = |E_{k,l}\rangle$ with an eigenvalue $\xi=\exp(-i E_k \tau)$. Physically this state corresponds to a dark state \cite{Plenio1998,Caruso2009,Thieldark}. Namely if such a state exists, and if it is chosen as the initial condition, it will never be detected. So, in this case, the condition of null measurements is guaranteed from the start. Similarly, it is easy to see that under the same condition, i.e., $\langle \psi_{\rm d}|E_{k,l}\rangle=0$, the corresponding left eigenvector is $\langle \xi^L| = \langle E_{k,l}|$. We see in this special example a property which is unique to states with eigenvalues on the unit circle, namely $|\xi^L\rangle=|\xi^R\rangle$. This feature is of some importance later. 

Our goal now is to consider other dark states, that are eigenstates of the survival operator, with eigenvalues on the unit circle. Recall that such states obviously have the largest possible eigenvalue and hence dominate the large-$n$ limit, if they exist. We use the energy representation and consider a degenerate subspace in the Hilbert space namely $\{ |E_{k,1}\rangle, \cdots, |E_{k,g_k}\rangle \}$ and by definition all these states have the same energy $E_k$. If some of them satisfy $\langle \psi_{\rm d}|E_{k,l}\rangle=0$ they are clearly dark as we have just explained. Consider now a linear combination of two energy states and we index these with $l=1$ and $l=2$. We then find the eigenvector of $\hat{\mathfrak{S}}$
\begin{equation}
	|\xi^R\rangle=N(\langle \psi_{\rm d}|E_{k,2}\rangle| E_{k,1}\rangle - \langle \psi_{\rm d}|E_{k,1}\rangle| E_{k,2}\rangle).
	\label{eq:two dark}
\end{equation}
Here clearly $\hat{\mathfrak{S}}|\xi^R\rangle=\exp(-i E_k \tau)|\xi^R\rangle$ since $U|\xi^R\rangle = \exp(-i E_k \tau)|\xi^R\rangle$ and $D |\xi^R\rangle=0$, and so the eigenvalue of the survival operator is $\xi=\exp(-i E_k \tau)$, namely it is just a phase determined by the energy $E_k$. Similarly, it is easy to see that $\langle \xi^L|=\langle \xi^R|$ is a left eigenvector of the survival operator, and hence $\langle \xi^L|\xi^R\rangle=1$. So these vectors are parallel, unlike the left and right vectors we found above with zero eigenvalue $\xi=0$ (and unlike the eigenvectors we discuss below with $|\xi| \textless 1$). The state in Eq. (\ref{eq:two dark}) is clearly dark in the sense that if we start in this state, the amplitude on the detected state is always zero. This, in turn, is because this is a stationary state of $H$, hence between measurements, the dynamics do not allow the leakage of probability amplitude to the detected state, so we never detect the system. Once again, by a dark state, we mean that even if we do not condition the measurements to be null, the sequence of measurements will never detect the particle.

We can easily construct other normalized dark states using similar methods. We denote these states, i.e. states that cannot be detected as $|\delta_{k,m}\rangle$ ($\delta$ is for dark). Following \cite{Thieldark}, we find that in an energy subspace with $g_k$ states, there are $g_k -1$ dark states $\{|\delta_{k,m}\rangle\}$ with $ m=1, \cdots, g_k-1$. These states are given by a Gram Schmidt procedure \cite{Thieldark,PhysRevLett.89.080401,Facchi_2008,Caruso2009,doi:10.1002/andp.201600206}
\begin{equation}
    |\delta_{k,m}\rangle=N\sum_{j=1}^m \Bigg[|\alpha_{k,j}|^2 |E_{k,m+1}\rangle-\alpha_{k,m+1}^* \alpha_{k,j} |E_{k,j}\rangle\Bigg],
    \label{darkstateeq}
\end{equation}
where $\alpha_{k,m}=\langle E_{k,m}| \psi_{\rm d} \rangle$ and $m$ goes from $1$ to $g_k-1$. Here   $|E_k -E_i|\tau \neq 2 \pi j$, where $j$ is an integer. All these states are orthonormal with respect to each other, and again they are stationary states of the system, namely $H |\delta_{k,m}\rangle = E_k |\delta_{k,m}\rangle$. It is easy to see that these states are eigenstates of the survival operator with eigenvalue $\xi=\exp(-i E_k \tau)$, namely these eigenvalues all fall on the unit circle, since $\hat{\mathfrak{S}}|\delta_{k,m}\rangle= \exp(-i E_k \tau)|\delta_{k,m}\rangle$. It is also easy to show that the left eigenvectors satisfy $\langle \xi^L|=\langle \xi^R|$.

We have seen how dark states can be presented as energy eigenstates of $H$ but also as eigenstates of the survival operator $\hat{\mathfrak{S}}$. The moduli of the corresponding eigenvalues of $\hat{\mathfrak{S}}$ are unity.  As mentioned, with a $g_k$-fold degenerate energy subspace $\{|E_{k,1}\rangle,\cdots,|E_{k,g_k}\rangle \}$, we can generate $g_k-1$ dark states. In this subspace, there exists one additional state, which we call the bright state, and it is given by \cite{Thieldark}
\begin{equation}
	|\beta_k\rangle =N \hat{P}_k |\psi_{\rm d}\rangle,
	\label{bright}
\end{equation} 
where  $\hat{P}_k := \sum_{m=1}^{g_k} |E_{k,m}\rangle\langle E_{k,m}|$ is the eigenspace projector. This state is not an eigenvector of the survival operator. Its mathematical property is that it is orthogonal to all the $g_k -1$ dark states in its energy sector. This state, under repeated measurements, without conditioning the outcome, will be eventually detected with probability one; hence it is called a bright state. In the next section, we will show how to use these bright states to construct the eigenstates of the survival operator $\hat{\mathfrak{S}}$  with eigenvalues in the unit disk, i.e., $ 0\textless |\xi|\textless 1$.

To conclude this part, we see that the operator $\hat{\mathfrak{S}}$ has an eigenvalue zero and we have formally found the normalized left and right eigenvectors which are not parallel $\langle \xi^L|\xi^R\rangle \neq 1$. Using the energy presentation, we see that dark states of the system are eigenvectors of the survival operator, with eigenvalue $\xi =\exp(-i E_k \tau)$. Clearly $|\xi|=1$, so these have the maximum possible eigenvalue magnitude, since in general $|\xi| \leq 1$. In this case the left and right eigenvalues are parallel $ \langle \xi^L|\xi^R\rangle=1$. In a degenerate sector $\{|E_{k,l}\rangle\}$ of the Hilbert space, we have $g_k -1$ such eigenvectors that are given in Eq. (\ref{darkstateeq}) and they all share the same eigenvalue $\exp(-i E_k \tau)$.

For example, consider a system with three distinct energy levels, the first with degeneracy 10, the second with degeneracy 5, and another with no degeneracy. Further assume for simplicity that all the energy states have finite amplitude on the detected state. Eq. (\ref{darkstateeq}) specifies 13 orthonormal eigenstates of the survival operator, all with eigenvalues on the unit circle. These are dark states as mentioned, namely, states that cannot be detected even if we do not condition the measurements. However, if we consider the case where we choose the detected state as the energy eigenstate of the non-degenerate level, it then immediately follows from standard quantum mechanics that the remaining 15 states are all dark as they are orthogonal to the detected state. Hence, these 15 states are dark, and they are also eigenvectors of the survival operator $\hat{\mathfrak{S}}$.

If the system has no degeneracy, for example, a typical random system with no symmetry in $H$, and if all the energy states have finite overlaps with the detected one (even small), we reach the conclusion that we have no dark states, and hence we do not have eigenstates of the survival operator with eigenvalues $|\xi|=1$. This indicates that the dynamics of systems with disorder and those without, under the condition of repeated null measurements, can have very different properties. 

There exists, of course, a third family of states, which are those with $ 0 \textless |\xi| \textless 1$, namely the eigenvalues are all in the unit disk. There is no simple way to determine the values of $\xi$ and find the corresponding states. However, there is an elegant method to find these eigenvalues in principle based on a classical charge theory \cite{Gruenbaum2013}, which is discussed below. With this classical theory, we will be able to gain physical insight on the largest eigenvalues of systems with no dark states, which in turn will give the long-time dynamics of the conditioned measurement process.

\section{Classical charge picture mapping and biorthogonal eigenstates}\label{sec:charge bio}
\begin{figure*}
    \centering
    \includegraphics[width=0.8\linewidth]{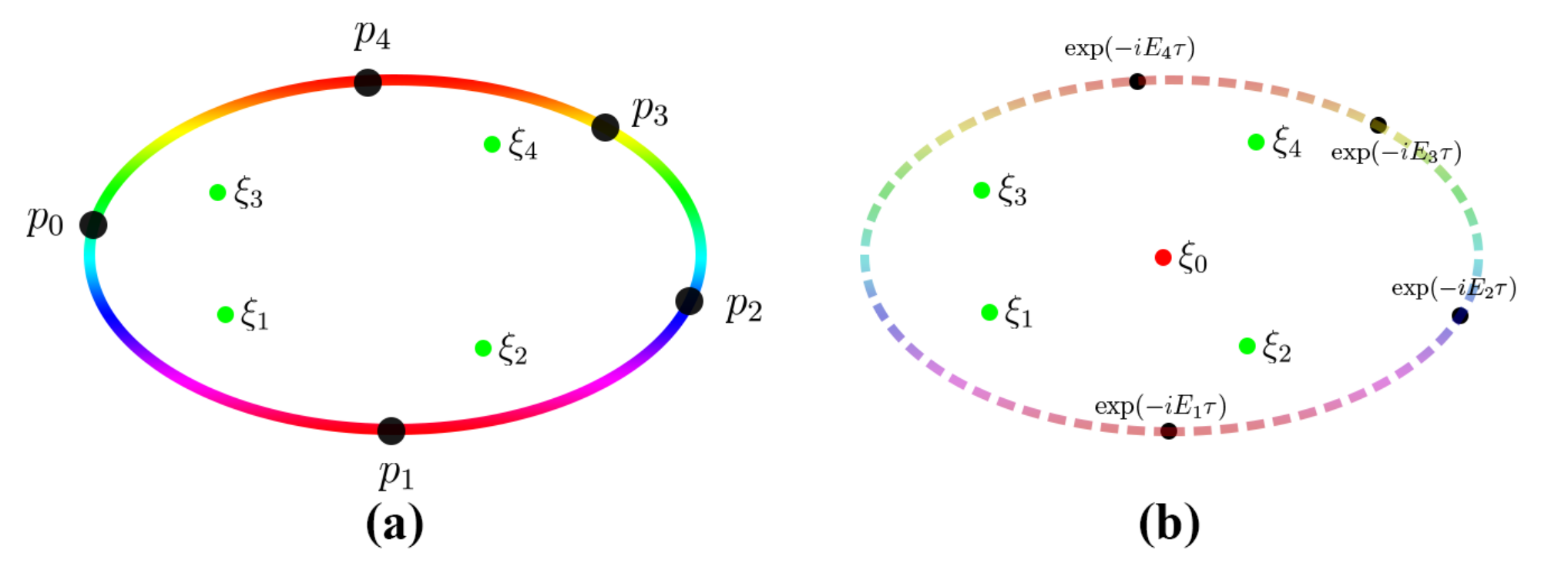}
    \caption{Schematic depiction of the classical charge mapping \cite{Gruenbaum2013} $(a)$ and the eigenvalues of survival operator $\hat{\mathfrak{S}}$ $(b)$. We plot a system with energy levels $ E_0, \cdots, E_4$, where all the energy levels are two-fold degenerate, except for $E_0$ which has not degeneracy. According the the charge picture, we have five charges on the unit circle, as shown on the left, the charges are $p_0, \cdots, p_4$. These charges are located on $\exp(- i E_k \tau)$ from Eq. (\ref{charge}). All these charges are positive, and hence this gives 4 stationary points in the unit disk denoted $\xi_1, \cdots, \xi_4$ in $(a)$. These, as explained in the text, give the non-zero eigenvalues of the survival operator in the unit disk $0\textless |\xi_i| \textless 1$. In $(b)$, we plot all the eigenvalues of $\hat{\mathfrak{S}}$. There are three types of eigenvalues, i.e., $|\xi|=1$, $0\textless |\xi|\textless 1$, and $\xi_0=0$.  The degeneracy of the energy levels leads to the dark states, which are eigenstates of $\hat{\mathfrak{S}}$ with eigenvalues $\exp(- i E_k \tau)$ on the unit circle $(b)$. The eigenvalues in the unit circle are given by the charge picture, and finally there is one eigenvalue that is at the center of the unit disk. Note that the non-degenerate energy level $E_0$ does not contribute a dark state. }
    \label{fig:map}
\end{figure*}

As previously mentioned, the dark states are eigenfunctions of survival operator with a special property; their eigenvalues are on the unit circle. However, the dark states are certainly not the only part of the eigenstates of the survival operator, namely, the dark space ${\cal H}_D:=\text{Span}(\{|\delta_{k,m}\rangle\})$ is a subspace of the full Hilbert space ${\cal H}$, so we need to search for other eigenstates. The eigenvalues of the remaining states, which are now investigated, all fall in the unit disk. Using the matrix determinant lemma \cite{harville2006matrix}, the eigenvalues of the survival operator $  \hat{\mathfrak{S}}$ are given by (see Appendix \ref{ap:A})
\begin{equation}
	\text{det}[\xi \mathbb{1} -\hat{\mathfrak{S}}]=\text{det}[\xi \mathbb{1} - \hat{U}(\tau)] \langle \psi_{\rm d} | [\xi \mathbb{1} - \hat{U}(\tau)]^{-1} |\psi_{\rm d}\rangle \xi = 0.
	\label{eq:det}
\end{equation} 
We observe the following:

(i) There is a stationary point, denoted $\xi_0$, located at the origin, namely $\xi_0=0$ (see the last term in Eq. (\ref{eq:det})). We have found its left and right eigenvectors in the former section.

(ii) For simplicity, let us assume all elements of the energy basis $\{|E_{k,l}\rangle\}$ have finite overlaps with $|\psi_{\rm d}\rangle$. Using this energy basis, we find:
\begin{equation}
	\text{det}[\xi \mathbb{1} - U(\tau)] =\Pi_{k=0}^{w-1} (\xi-e^{-i E_k \tau})^{g_k}.
	\label{eq:det1}
\end{equation}
We have here a multiplication of $\varpi$ terms $(\xi-e^{-i E_k \tau})$, where $\varpi$ is the size of the full Hilbert space and $w$ is the number of distinct energy levels, namely $\sum_{k=0}^{w-1} g_k =\varpi$. Only in the absence of degeneracy, $g_k=1$, we have $\varpi=w$. Here following Gr\"{u}nbaum \textit{et al.} \cite{Gruenbaum2013,Grunbaum2014}, we use the symbol $w$, as this is actually describing a certain winding number of the problem. Naively, it seems from Eq. (\ref{eq:det1}) that we already have all $\varpi$ $\xi$s that fit Eq. (\ref{eq:det}). However, this intuition is not true, because $\langle \psi_{\rm d} | [\xi \mathbb{1} - \hat{U}(\tau)]^{-1} |\psi_{\rm d}\rangle$ cancels part of them, as we now show.

(iii) The bright and dark space, i.e.,  $\{|\beta_i\rangle_{i=0}^{w-1} \}$ and $\{|\delta\rangle\}$, form a complete basis. Importantly $|\psi_{\rm d}\rangle$ is orthogonal to the dark states by definition. It follows that we can expand the detected state in terms of bright subspace only
 \begin{equation}
 	|\psi_{\rm d}\rangle = \sum_{k=0}^{w-1} \langle \beta_k|\psi_{\rm d}\rangle |\beta_k\rangle.
 	\label{eq:det psi}
 \end{equation}
 The bright states $\{|\beta_k\rangle_{l=0}^{w-1} \}$ are also eigenstates of $H$ and hence of $\hat{U}(\tau)$. It follows that
 \begin{equation}
 	\langle \psi_{\rm d}|[\xi \mathbb{1}-\hat{U}(\tau)]^{-1}|\psi_{\rm d}\rangle = \sum_{k=0}^{w-1} \frac{| \langle \beta_k|\psi_{\rm d}\rangle|^2 }{\xi- e^{-i E_k \tau} }.
 	\label{eq:det2}
 \end{equation}
 Using Eqs. (\ref{eq:det}, \ref{eq:det1}, \ref{eq:det psi}, \ref{eq:det2}) , the equation for $\xi$s reads:
 \begin{equation}
 	\sum_{k=0}^{w-1} |\langle \beta_k|\psi_{\rm d}\rangle|^2 (\xi- e^{-i E_k \tau})^{g_k-1} \Pi_{i \neq k} (\xi-e^{-i E_i \tau})^{g_i}=0.
 \end{equation}
From here we see that if an energy level is degenerate $g_k \textgreater 1$, then we find the survival operator $\hat{\mathfrak{S}}$ has eigenvalues $\xi= \exp(- i E_k \tau)$, which is $(g_k -1)$-fold degenerate. In contrast, if $g_k=1$, $\xi= \exp(-i E_k \tau)$ is clearly not an eigenvalue of $\hat{\mathfrak{S}}$. This  coincides with what we have found in the previous section---any degenerate subspace with energy $E_k$ has $g_k-1$ dark states and one bright state. The $(g_k -1)$-fold eigenvalues of the survival operator  $\xi= \exp(- i E_k \tau)$ correspond to $g_k-1$ dark states found in Eq. (\ref{darkstateeq}). The total number of dark states is $\sum_{k=0}^{w-1} (g_k-1)=\varpi - w$. These states, have the largest possible eigenvalue magnitude $|\xi|=1$ of $\hat{\mathfrak{S}}$, hence they will control the long-time dynamics of the conditional measurements process, see below. For disordered, interacting, or chaotic systems, where repulsion of energy levels arises, namely $g_k=1$ for all the energy levels, we cannot find eigenvalues on the unit circle in general. Now $\varpi =w$ and $\sum_{k=0}^{w-1} (g_k-1)=0$.

(iv) The other eigenvalues $0 \textless |\xi_{\iota} | \textless 1 $, are given by
\begin{equation}
	\langle \psi_{\rm d} | [\xi \mathbb{1} - \hat{U}(\tau)]^{-1} |\psi_{\rm d}\rangle=0.
	\label{eq:15}
\end{equation}
It turns out that there are $w-1$ $\xi$s, that satisfy $0 \textless |\xi_{\iota} | \textless 1 $. So we see, we have $\varpi - w$ eigenvalues on the unit circle $|\xi|=1$, $w-1$ in the unit disk $0\textless |\xi|\textless 1$, and one with $\xi=0$.  $(\varpi - w) +(w-1)+1=\varpi$, all the eigenvalues of survival operator $\hat{\mathfrak{S}}$ are found.

 The question remains, how to find the eigenvalues satisfying $0 \textless |\xi_{\iota} | \textless 1 $?  Here we exploit a beautiful mapping of the problem to a classical charge theory, following the work of Gr\"{u}nbaum \textit{et al.} \cite{Gruenbaum2013,Yin2019,PhysRevResearch.2.033113}. By defining $p_k= \langle \psi_{\rm d} | \hat{P}_k |\psi_{\rm d}\rangle=\sum_{i=1}^{g_k} | \langle E_{k,i}|\psi_{\rm d}\rangle|$, together with Eqs. (\ref{bright}), (\ref{eq:det2}) and (\ref{eq:15}), we get
\begin{equation}
    \mathcal{F}(\xi)=\sum_{k=0}^{w-1}\frac{p_k}{\xi-e^{-i E_k \tau}}=0.
    \label{charge}
\end{equation}
$\mathcal{F}(\xi)$ can be considered as a  two-dimensional electrostatic field created by point charges $p_k$ at positions $e^{-i E_k \tau}$ (see Fig. \ref{fig:map}). These charges produce a $\log$ potential, namely they can be viewed as long wires piercing the unit disk. Eq. (\ref{charge}) defines the stationary points of this classical field. Namely, when we put a test charge into this field, the stationary points $\{\xi_{\iota} \}$ are the locations where the net force on the test charge is zero.   All the $\{\xi_{\iota} \}$ are inside the unit disk ($|\xi_{\iota} |<1$), which is rather obvious since all the charges $p_k$ are positive. 

To summarize to find the eigenvalues according to the classical charge picture, we consider: i) $w$ charges that are placed on the unit circle, where $w$ is the number of distinct energy levels of $H$; ii) these charges are located at the phases $\exp(- i E_k \tau)$, where $E_k$ are the energy levels of $H$; (iv) the charges have magnitude $p_k$; (v) with these positive charges, we have $w-1$ zeros of the force field, all inside the unit disk; (vii) once we calculate these zeros, these are the eigenvalues $0 \textless |\xi| \textless 1 $ we are looking for. The charge theory is useful as it allows us to easily identify the largest eigenvalue of the survival operator, in certain limits of the problem discussed below.

\subsection{Eigenvectors of the survival operator}

Now we formally find the eigenvectors of $\hat{\mathfrak{S}}$ that correspond to the eigenvalues that are in the unit disk.  Here we take advantage that the dark states are already eigenstates of $\hat{\mathfrak{S}}$ that correspond to eigenvalues on the unit circle. Furthermore, as mentioned, the dark and bright subspace form a complete basis. With these two preconditions, we expand $|\xi^R\rangle$ in terms of the bright states, under the condition that the eigenvalue of $|\xi^R\rangle$ is in the unit disk. As presented in Appendix \ref{calculate the xi}, we find:
\begin{equation}
	|\xi_{\iota} ^R\rangle= N\sum_{j=0}^{w-1} \hat{R}_{\iota ,j}^{\dag} |\psi_{\rm d}\rangle, \quad \hat{R}_{\iota ,j}^{\dag} = \frac{ \hat{P}_j}{\xi_{\iota} -e^{-i E_j \tau}}.
	\label{righte}
\end{equation}
Using Eq. (\ref{bright}), we see that the states $\{|\xi_{\iota}^R\rangle\}$ indeed are linear combinations of the bright states $\{|\beta_k\rangle\}$ as stated. So $|\xi^R\rangle$ is a bright state as well, namely in the absences of conditioning, starting in this state, the detection (a click yes) is eventually guaranteed. Though both $\{|\beta_k\rangle\}$ and $\{|\xi_{\iota}^R \rangle\}$ are bright states, the latter are eigenstates of the survival operator $ \hat{\mathfrak{S}} $ with corresponding eigenvalues $\{\xi_{\iota}\}$ in the unit disk, while the former are eigenstates of $H$ and $\hat{U}(\tau)$. Note that the right eigenvectors of $ \hat{\mathfrak{S}} $ do not have to be orthogonal with each other, i.e., $\langle \xi_{\iota}^R| \xi_{\iota^{\prime}}^R\rangle \neq \delta_{\iota ,\iota^{\prime} }$. 

In Eq. (\ref{righte}), the $\{|\xi_{\iota} ^R\rangle\}$ are presented in an energy basis, in which the evolution operator $\hat{U}(\tau)$ is diagonal. Using the identity $\sum_{j=0}^{w-1} \hat{P}_j =1$, we have $[\xi_{\iota} \mathbb{1} -\hat{U}(\tau)]^{-1}= \sum_{j, j^{\prime}=0}^{w-1} \hat{P}_j [\xi_{\iota} -\hat{U}(\tau)]^{-1} \hat{P}_{j^{\prime}}= \sum_{j=0}^{w-1} \hat{P}_j/(\xi_{\iota}-e^{- i E_j \tau}) = \sum_{j=0}^{w-1} \hat{R}^{\dag}_{\iota,j} $. So
\begin{equation}
	|\xi_{\iota} ^R\rangle = N [\xi_{\iota} \mathbb{1} -\hat{U}(\tau)]^{-1}|\psi_{\rm d}\rangle,
	\label{eq:rightG}
\end{equation}
independent of any representation. The geometry of the right eigenstates is that they are all orthogonal with respect to the detection state. This is easy to see using Eq. (\ref{eq:1}), and $\langle \psi_{\rm d}|(\mathbb{1}-D)=0$.  

Here we present a direct proof to see that Eq. (\ref{eq:rightG}) is indeed the right eigenvector of $ \hat{\mathfrak{S}} $. We start from the definition of the right eigenvectors. Using Eq. (\ref{eq:eigenG}), we have:
\begin{equation}
	(\hat{U}-\xi_{\iota} \mathbb{1} )|\xi_{\iota}^R\rangle=|\psi_{\rm d}\rangle \langle \psi_{\rm d}|\hat{U}|\xi_{\iota}^R\rangle.
	\label{eq:19}
\end{equation}
Inserting Eq. (\ref{eq:rightG}) into Eq. (\ref{eq:19}), we get
\begin{equation}
	-|\psi_{\rm d}\rangle =|\psi_{\rm d}\rangle \langle \psi_{\rm d}| \frac{\hat{U}}{\xi_{\iota} \mathbb{1}-\hat{U}}|\psi_{\rm d}\rangle,
\end{equation}
which leads to $-1=\langle \psi_{\rm d}|\hat{U}/(\xi_{\iota} \mathbb{1}-\hat{U})|\psi_{\rm d}\rangle$. Using $\hat{U}/(\xi_{\iota} \mathbb{1} -\hat{U})+1=\xi_{\iota} \mathbb{1} /(\xi_{\iota} \mathbb{1} -\hat{U})$, we get for $\xi_{\iota} \neq 0$
\begin{equation}
	\langle \psi_{\rm d}|[\xi_{\iota} \mathbb{1} -\hat{U}]^{-1}|\psi_{\rm d}\rangle =0.
	\label{eq:p21}
\end{equation}
From Eq. (\ref{eq:15}), this is the equation for the eigenvalues of the survival operator $ \hat{\mathfrak{S}} $ that are in the unit disk, i.e., $0\textless |\xi|\textless 1$. Hence, Eq. (\ref{eq:rightG}) is correct.

  Following the same procedure, the left eigenvectors of survival operator read:
\begin{equation}
	\langle \xi_{\iota} ^L|= N\sum_{j=0}^{w-1} \langle \psi_{\rm d}| L_{\iota,j}, \quad L_{\iota ,j}= \frac{\hat{P}_j e^{-i E_j \tau}}{\xi_{\iota} - e^{- i E_j \tau}}.
	\label{eq:lefteigen}
\end{equation}
Similar to Eq. (\ref{eq:rightG}), we have
\begin{equation}
	\langle \xi_{\iota} ^L|=N\langle \psi_{\rm d}| \hat{U}(\tau)[\xi_{\iota} \mathbb{1} - \hat{U}(\tau)]^{-1} .
	\label{eq:leftG}
\end{equation}
The right/left eigenvectors obey the biorthogonal  relation $\langle \xi_{\iota} ^L | \xi_{\iota^{\prime}} ^R\rangle = N_{\iota} \delta_{\iota ,\iota^{\prime} }$. Here $N_{\iota}$ is a constant which depends on how we normalize the two right/left vectors. Furthermore, the right/left eigenvectors are orthogonal to the dark states $\langle \delta_{k,j} | \xi_{\iota} ^R\rangle=0$ and $\langle \xi_{\iota} ^L | \delta_{k,j} \rangle=0$, since the dark subspace is orthogonal to the bright subspace.

\subsection{Phase  gained under repeated measurements}

We now explain the physical meaning of $\xi$.  Using Eqs. (\ref{eq:rightG}) and (\ref{eq:leftG}), if the system starts with the normalized right eigenstate $|\xi_{\iota}^R\rangle$, i.e, $|\psi_{\rm in}\rangle =|\xi_{\iota}^R\rangle$, we have (Appendix \ref{apd:relation})
\begin{equation}
	|\xi_{\iota}^R\rangle \stackrel{\hat{U}(\tau)}{\longrightarrow} |\xi_{\iota}^L\rangle^*  \stackrel{(\mathbb{1}-D)}{\longrightarrow} |\xi_{\iota}^R\rangle, \quad |\xi_{\iota}^R\rangle \stackrel{\hat{\mathfrak{S}}}{\longrightarrow} |\xi_{\iota}^R\rangle, 
	\label{eq:jump}
\end{equation} 
here $*$ is the complex conjugate. And we use the fact that  $|\xi_{\iota}^R\rangle$ is an eigenstate of $\hat{\mathfrak{S}}$, so $\hat{\mathfrak{S}}|\xi_{\iota}^R\rangle \rightarrow |\xi_{\iota}^R\rangle$.  Using Eq. (\ref{eq:waveG}), the wave function of the system after $n$ measurements reads:
\begin{equation}
	|\psi_n\rangle =N_n \xi_{\iota}^n|\xi_{\iota}^R\rangle,
\end{equation} 
where $N_n$ is normalization. In this process, the operations $\hat{U}(\tau)$ and $(\mathbb{1}-D)$ send the wave function to the $|\xi_{\iota}^L\rangle^*$ and  $|\xi_{\iota}^R\rangle$ $n$ times. Importantly, we gain an unusual phase, namely
\begin{equation}
\boxed{
	|\psi_n\rangle = \exp(i n \phi)|\xi_{\iota}^R\rangle= \exp(i n \phi)|\psi_{\rm in}\rangle},
	\label{eq:21}
\end{equation}
where $\phi= -i \ln(\xi_{\iota}/|\xi_{\iota}|)$. This phase is not determined by the energy times evolution time, but is given by the phase of the complex eigenvalue $\xi_{\iota}$ of the survival operator $\hat{\mathfrak{S}}$.  As we repeat the measurements, the system is driven periodically, and the phase accumulated is $n\phi$.

\subsection{Exceptional points: example of a two-level system}\label{sec:two-level}
 In the above discussion we assumed that the system has no exceptional points. Exceptional points are cases, where for specific $\tau$, or for certain  control parameters of $H$, we have two (or more) eigenvalues $\xi$ ($|\xi|\textless 1$) of the survival operator $\hat{\mathfrak{S}}$ merging, a kind of degeneracy found  for non-Hermitian physical systems \cite{Mirieaar7709}. In this subsection, we illustrate part of the effects with the simplest example---the two-level system, also presenting the exceptional point and its meaning. More general discussion about the exceptional points is presented in Sec. \ref{sec:ep}.

 For a two-level system with state $|l\rangle$ (left) and $|r\rangle$ (right), the Hamiltonian reads $H=-\gamma (|l \rangle \langle r|+|r \rangle \langle l|)$  where $\gamma$ is the hopping amplitude between these two states. The energies in this system are $E_1 = -\gamma$ and $E_2 = \gamma$. We assume that we detect the particle on the left node, so $D= | l \rangle \langle l |$. In this basis the survival operator is
\begin{equation}
	\hat{\mathfrak{S}}= (\mathbb{1}-D)\hat{U}(\tau)=\left(
	\begin{matrix}
		0 & 0\\	
		i \sin(\gamma \tau) & \cos(\gamma \tau)
	\end{matrix}\right).
	\label{eq:two-level}
\end{equation}
We easily find the following eigenvalues and vectors of $\hat{\mathfrak{S}}$
\begin{equation}
	\begin{aligned}
		&\xi=0: \qquad \ \ \langle \xi^L|= \langle l |, \quad |\xi^R\rangle = \cos(\gamma \tau)|l \rangle - i \sin(\gamma \tau) |r \rangle;\\
		&\xi=\cos(\gamma \tau): \langle \xi^L| = i \sin(\gamma \tau)\langle l |+\cos (\gamma \tau)\langle r|, \ \  |\xi^R\rangle = |r\rangle.
	\end{aligned}
	\label{eq:two-level eigen}
\end{equation}
We see from here the features discussed above, for example for the eigenvalue $\xi=0$ the left eigenstate is $\langle \xi^L|=\langle \psi_{\rm d}|$, which in this model is $\langle \xi^L|=\langle l|$. Further
\begin{equation}
	\hat{U}(\tau)=\left(
	\begin{matrix}
		\cos(\gamma \tau) & i \sin(\gamma \tau)\\
		i \sin(\gamma \tau) & \cos(\gamma \tau)
	\end{matrix}\right),
\end{equation}
and hence for the second eigenvalue $\xi=\cos(\gamma \tau)$ and vector $|\xi^R\rangle =|r\rangle$, we have $\hat{U}(\tau)|\xi^R\rangle=i \sin(\gamma \tau)|l \rangle +\cos(\gamma \tau)|r\rangle=|\xi^L\rangle^*$. Then the conditional measurement $(\mathbb{1}-D)$ wipes the amplitude on $|l\rangle$, which leads to $(\mathbb{1}-D)|\xi^L\rangle^* \rightarrow |r\rangle=|\xi^R \rangle $. This is an illustration of Eq. (\ref{eq:jump}).

 According to our formalism, in the large-$n$ limit the state of the system is determined by the eigenvector corresponding to the eigenvalue closest to the unit circle, i.e. $\text{max}(0, |\cos(\gamma \tau)|)$ which, unless $|\cos(\gamma \tau)|=0$, means $\lim_{n \rightarrow \infty} |\psi_n \rangle = |r \rangle$. This is easy to understand, as after the null measurements, the system cannot be on $|l \rangle$. Since there are only two states, we find the system after the $n$-th measurement on $|r \rangle$. This holds true for any $n$.

The exceptional point is found when $\cos(\gamma \tau)=0$, and for this case $\hat{\mathfrak{S}}$ cannot be diagonalized at all. Three features are found: a) in this limit, Eq. (\ref{eq:two-level eigen}) gives $\langle \xi^L|\xi^R\rangle =0$, b) the two eigenvalues coalesce on zero, i.e. $\xi=0$, finally c) the two pairs of vectors, both $L$ and $R$, in Eq. (\ref{eq:two-level eigen}) become parallel (in our representation besides an $i$ they are identical). Thus at the exceptional sampling time $\tau= (\pi/2+k \pi)/\gamma$, where $k$ is an integer, we have only one eigenvector instead of the familiar two. This means that both the eigenvalues and the eigenvectors coalesce, which implies an exceptional point. Mathematically, when $\cos(\gamma \tau)=0$, the survival operator now reads
\begin{equation}
	\hat{\mathfrak{S}}=(\mathbb{1}-D)\hat{U}(\tau)=
	\left(
	\begin{matrix}
		0 & 0\\
		i & 0
	\end{matrix}\right)
\end{equation}
and then the eigenvalues are clearly zero, and we have only one eigenvector solution instead of two.

How can we search for these exceptional points without solving the problem exactly? We claim that we can use the classical charge picture, which is very useful in determining the exceptional points in general.  Let us present it for the two-level system so that the reader can get familiarized with the basic ideas. We allocate the classical charges on the unit circle. The location of the pair of charges, are on the phases $\exp(- i E_1 \tau)= \exp(i \gamma \tau)$ and $\exp(- i E_2 \tau)= \exp(-i \gamma \tau)$. The magnitudes of the charges in our example are identical and positive. Then we search for the stable point, namely the point in space where the classical force field of the charges vanishes. This is clearly on the midpoint connecting two charges. From basic geometry, we realize that this classical stable point is found in this example on $\xi=\cos(\gamma \tau)$ in the complex plane. Especially, if we choose $\tau$, such that these charges are on the north and south poles, the stable point is on the origin, in agreement with Eq. (\ref{eq:two-level eigen}). This describes the coalescence at the stationary point. Dynamically we imagine the eigenvalue $\cos(\gamma \tau)$, moving to the origin where it fuses with the ever-present eigenvalue $\xi=0$ there. By moving, we mean that we think of the process as we tune $\gamma \tau$. We may control this fusing process with the rearrangement of the charges on the unit circle, i.e., moving them to the south and north pole.  The mapping of the problem to a classical charge theory allows for an intuitive understanding when exceptional points emerge, in generality, and shows how this is related to the symmetry of the problem.

For the two-level system, what is the physical meaning of the exceptional point? When the system is tuned to this point, and we attempt to follow the null measurements, we encounter a problem. After the first measurement, we get a null result, as this is the condition imposed by the rules of the model. But now, after the first measurement, the system is in state $|r \rangle$ (since we conditioned the measurements to give a no when we detect on $|l \rangle$). However, when $\cos(\gamma \tau)=0$, we will find the state of the system just before the second measurement in the state $|l \rangle$. We then perform the second measurement and find the system in the detected state $|l\rangle$ with probability one. It follows that we tried to impose the sequence of null measurements. However, at the exceptional points, this is impossible since we always get a yes in the second measurement. This feature is generic, and found below in systems whose complexity exceeds that of a two-level example.

\begin{widetext}

\section{State function under conditioned measurements}\label{sec:formal}
A useful identity, which holds for non-Hermitian operators, and not limited to our case reads \cite{datta2016matrix,Brody_2013,Ghatak2019}
\begin{equation}
	\sum_{\xi} \frac{|\xi^R \rangle \langle \xi^L|}{\langle \xi^L| \xi^R\rangle }=1.
	\label{eq:identity}
\end{equation}
This formula is valid provided that we do not have exceptional points in the system. The summation is over all the states in the system, denoted by the sum over the index $\xi$.  It follows from Eq. (\ref{eq:identity}) that a generic initial condition can be expanded like
\begin{equation}
	|\psi_{\rm in}\rangle = \sum_{\xi} \frac{\langle \xi^L| \psi_{\rm in}\rangle |\xi^R\rangle}{\langle \xi^L|\xi^R\rangle}.
\end{equation}
For the case under study, we split the sum into three parts, those with eigenvalues on the unit circle $|\xi|=1$, $\xi=0$, and all the rest.
\begin{equation}
	|\psi_{\rm in}\rangle = \underbrace{ \frac{\langle \psi_{\rm d}|\psi_{\rm in} \rangle }{\langle \psi_{\rm d}| U^{-1}|\psi_{\rm d} \rangle } U^{-1} |\psi_{\rm d}\rangle}_{\xi=0}+ \sum_{ 0 \textless |\xi|\textless 1} \frac{ \langle \xi^L|\psi_{\rm in}\rangle}{\langle \xi^L|\xi^R\rangle }|\xi^R\rangle+ \sum_{\delta, |\xi|=1} \langle \delta |\psi_{\rm in}\rangle |\delta\rangle.
	\label{eq:wave ini}
\end{equation}
The first term, proportional to $U(\tau)^{-1}|\psi_{\rm d}\rangle$, has a clear physical meaning, since after the first time interval $\tau$, the unitary evolution yields   $UU^{-1} |\psi_{\rm d}\rangle = |\psi_{\rm d}\rangle$. The first detection, is modeled with the projector $(\mathbb{1}-D)$ and hence this first term is wiped out by the first measurement.   As mentioned, if $|\psi_{\rm in}\rangle = \hat{U}^{-1}|\psi_{\rm d}\rangle$, we cannot get the conditional measurements, namely for this initial condition, we record a yes at the first measurement attempt, so it is important for our discussion, that the second and third terms are non-zero. In Eq. (\ref{eq:wave ini}), we assume $\langle \psi_{\rm d}| U^{-1}|\psi_{\rm d} \rangle \neq 0$ and $\langle \xi^L|\xi^R\rangle \neq 0$, otherwise we get an exceptional point. Using Eqs. (\ref{eq:waveG}), (\ref{eq:eigenG}) and (\ref{eq:wave ini}), the wave function after $n \geq 1$ measurements reads
\begin{equation}
\boxed{
	|\psi_n\rangle = N \left\{  \sum_{ 0 \textless |\xi|\textless 1} \xi^n \frac{ \langle \xi^L|\psi_{\rm in}\rangle}{\langle \xi^L|\xi^R\rangle }|\xi^R\rangle+\sum_{\delta, |\xi|=1} \exp(- i E_{\delta} n \tau) \langle \delta |\psi_{\rm in}\rangle |\delta\rangle \right\}},
	\label{final}
\end{equation}
and $N$ is the normalization. In the next section we use this equation to explore the large-$n$ limit of the system. Here we see that in general the state of the wave function is composed of dark states $|\delta\rangle$ with eigenvalues $\exp(-i E_{\delta} \tau)$, and bright states $|\xi^R\rangle$ with eigenvalues $0<|\xi|<1$. The effect of the state with eigenvalue $\xi=0$ is washed out after the first measurement.

We now consider the mean energy of the system. Even though the particle is not detected, there must be an exchange of energy between the system and the detector, hence the energy of the particle is not constant. Initially, the expectation of the energy of the system is ${\cal E}_0=\langle \psi_{\rm in}|H|\psi_{\rm in}\rangle$. With the measurements going on, the averaged energy of the system after the $n$-th measurement is $\langle {\cal E}_n \rangle = \langle \psi_n | H |\psi_n\rangle$. Using Eq. (\ref{final}), we have:
\begin{equation}
	\langle {\cal E}_n \rangle = \frac{ \sum_{0\textless|{\xi},{\xi}^{\prime}|\textless 1 } a_{\xi}  a_{{\xi}^{\prime}}^* \xi^n (\xi^{\prime * })^n \langle \xi^{\prime R} | H | \xi^R\rangle +\sum_{\delta, |\xi|=1} |\langle \delta |\psi_{\rm in} \rangle |^2 E_{\delta} }{\sum_{0\textless|{\xi},{\xi}^{\prime}|\textless 1 } a_{\xi}  a_{{\xi}^{\prime}}^* \xi^n (\xi^{\prime * })^n \langle \xi^{\prime R} | \xi^R\rangle  +\sum_{\delta, |\xi|=1} |\langle \delta |\psi_{\rm in} \rangle |^2 },
	\label{eq:energyxi}
\end{equation}
where $a_{\xi}=\langle \xi^L | \psi_{\rm in}\rangle/\langle \xi^L | \xi^R\rangle$, and $\langle \widetilde{\xi}^{\prime R} | H | \xi^R\rangle = \sum_{k=0}^{w-1}p_k E_k /[(\widetilde{\xi}^{\prime} - e^{i E_k \tau})(\xi - e^{-i E_k \tau})]$.  If initially the system is prepared in a linear combination of dark states, then clearly the energy is a constant of motion, but otherwise it is not. In particular if we have no dark states in our system, namely no symmetry and hence no degeneracy,  the measurements never conserve energy. Note that in this case the mean energy can saturate for large $n$, or exhibit dynamical oscillations, which are explored below.

\end{widetext}

\section{The fixed state and quantum dynamics}\label{sec:fix and dynamics}
From Eq. (\ref{final}), in the large-$n$ limit, we reach several general conclusions:

$({ \cal A})$ If the dark subspace is not empty, namely if we have some degenerate energy levels (implying some symmetry in the system) or if one of the stationary states of $H$ is orthogonal to $|\psi_{\rm d}\rangle$, the second term in Eq. (\ref{final}) dominates the large-$n$ limit provided that the initial condition $|\psi_{\rm in}\rangle$ has some overlap with the dark states. This means that the long-time dynamics is determined by the phases $\exp(-i E_{\delta} n \tau)$. Compared to unitary dynamics, where all the energy levels contribute, the effect of measurement is therefore to remove the non-degenerate energy levels from the long-time dynamics. The wave function then reads:
	 \begin{equation}
	 \boxed{
		 |\psi_f\rangle = \lim_{n\rightarrow \infty}|\psi_n\rangle \sim N\sum_{\delta, |\xi|=1} e^{-i  E_{\delta} n \tau} \langle \delta| \psi_{\rm in} \rangle |\delta\rangle}.
		 \label{eq:dark only}
	 \end{equation}

$({ \cal B})$ In the case when the dark subspace is empty, as typically found for systems without special symmetries, we find two types of behaviors. We first need to determine the maximum of the set $\{|\xi_{\iota}|\}$ and recall that these absolute values are all less than unity. Then:

$({ \cal B}1)$ If the maximum is unique, the system in the long-time limit goes to a specific state.  There is no dynamics in this limit as only one term dominates. This is similar to a fixed point. We denote this unique largest eigenvalue of survival operator $\hat{\mathfrak{S}}$ as $\xi_f$.   Using Eq. (\ref{final}), we have
	   \begin{equation}
	   \boxed{
		|\psi_f\rangle \sim N \xi_f^n  |\xi_f^R\rangle = e^{i n \phi_f } |\xi_f^R\rangle},
		\label{eq:bright only}
	   \end{equation}
	where $\phi_f=-i\ln (\xi_f/|\xi_f|)$. The system ends up with the state $|\xi_f^R\rangle$ and attains a global phase $n \phi_f$.  In Eq. (\ref{eq:21}), the phase $\phi_i$ of the system is determined by the initial state, while here the phase $\phi_f$ is determined by the maximum of the set $\{|\xi_{\iota}|\}$. Furthermore, for a generic initial state, at the first stages of the processes ($n$ is not large) many phases  $\phi_i =-i\ln (\xi_i/|\xi_i|)$ contribute, see Eq. (\ref{final}). As the component $|\xi_f^R\rangle$  gradually wins the game, the corresponding phase becomes the only term that dominates ($n$ is large). Each application of $\hat{\mathfrak{S}}$ rotates the phase of the system by $\phi_f$.
	   
$({ \cal B}2)$ If the maximum of the set $\{|\xi_{\iota} |\}$ is shared by several eigenvalues of  the survival operator $\hat{\mathfrak{S}}$, all with the same absolute value but with different phases, then these eigenvalues will control the large-$n$ limit of Eq. (\ref{final}). In this case, the wave function, for large $n$, will exhibit dynamic. The system not only gains a global phase like the above case $({ \cal B}1)$, but also exhibits interference due to the relative phases of these eigenvalues.

As an example, consider two eigenvalues $|\xi_1|= |\xi_2| \textgreater |\xi_{i}| $ ($i \neq 1,2$) and $|\xi_1|=|\xi_2|<1$, we denote $\xi_1 = r e^{i \phi_1}$ and $\xi_2 = r e^{i \phi_2}$, where $r=|\xi_1|=|\xi_2|$, $\phi_1=-i \ln(\xi_1/|\xi_1|)$, and $\phi_2=-i \ln(\xi_2/|\xi_2|)$. Using Eq. (\ref{final}), the terms with eigenvalues $\xi_i$ ($i \neq 1,2$) decay away, and we are left with $|\psi_f\rangle= |\psi_{n}\rangle \sim N( a_1 r^n e^{i n \phi_1}   |\xi_{1}^R\rangle + a_2 r^n e^{i n \phi_2}  |\xi_{2}^R\rangle)= N r^n ( a_1  e^{i n \phi_1}   |\xi_{1}^R\rangle + a_2 e^{i n \phi_2}  |\xi_{2}^R\rangle)$ (as a reminder, $a_i$ is the overlap with the initial state). $r^n$ can be absorbed by the normalization. In the end, we have  
\begin{equation}
\boxed{
	|\psi_f \rangle \sim N e^{i n\frac{\phi_1+\phi_2}{2}} (a_1 e^{i n\frac{\phi_1-\phi_2}{2}} |\xi_1^R\rangle + a_2 e^{i n\frac{\phi_2-\phi_1}{2}} |\xi_2^R\rangle )}.
	\label{eq:quantum dynamics}
\end{equation}
Comparing with Eq. (\ref{eq:bright only}), when there are two $\xi$s that dominate the large $n$ evolution, the system not only gains a global phase $(\phi_1+\phi_2)n/2$ in the measurements process, but also exhibits an oscillation that is controlled by the relative phase $(\phi_1-\phi_2)n/2$ of $\xi_1$ and $\xi_2$. We can tune this relative phase by changing the sampling time interval $\tau$ (see Sec. \ref{Sec:charge}). Mean observables, like the energy, may thus exhibit periodic controllable oscillations, see our examples below. In this process, the system are steered periodically by the repeated measurements, and we call this interesting phenomena quantum dynamics induced by the measurements.

\section{Three-level system, $V$-shaped system}\label{Sec:three}
\begin{figure*}
    \centering
    \includegraphics[width=1\linewidth]{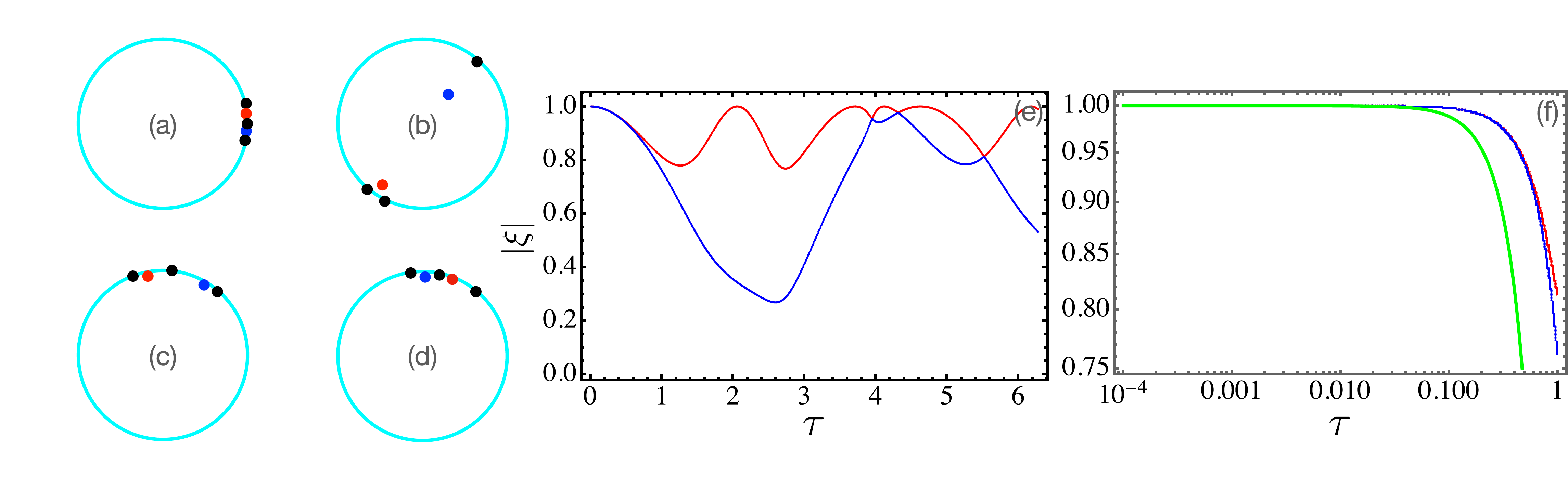}
    \caption{The absolute values of the eigenvalues $\xi_1$ (red) and $\xi_2$ (blue) versus $\tau$ are presented in sub-plots (e) and (f). Plots (a)-(d) are charge configurations for $\tau=0.1$ (a), $\tau=2$ (b), $\tau= 4.31697$ (c), and $\tau=4$ (d). The black points are charges located on the unit disk. The red and blue points are stationary points (eigenvalues of $\hat{\mathfrak{S}}$) $\xi_1$ and $\xi_2$. In this simple model, we have two eigenvalues $\xi_1$ and $\xi_2$  that are in the unit disk. The larger one, in absolute value sense, determines the final state of the system in the large-$n$ limit. When the curves in (e) cross, namely $|\xi_1|=|\xi_2|$, we find ever lasting quantum oscillations, for example for the energy.  In the Zeno limit $(a)$, $\tau \rightarrow 0$, both eigenvalues approach unity. A comparison with the lower bound  Eq. (\ref{eq:zeno cos}) (green line) is presented in (f). Note the usefulness of the charge picture. For example in (a), since $\tau$ is small, all the charges, i.e. black points on the unit disk, are closely situated, and it follows from basic electrostatics that also the eigenvalues (red and blue points) are both close to unity. This is shown clearly in subplot (e) in the small $\tau$ limit.  }
    \label{fig:threeabs}
\end{figure*}
\begin{figure*}
    \centering
    \includegraphics[width=2.1\columnwidth]{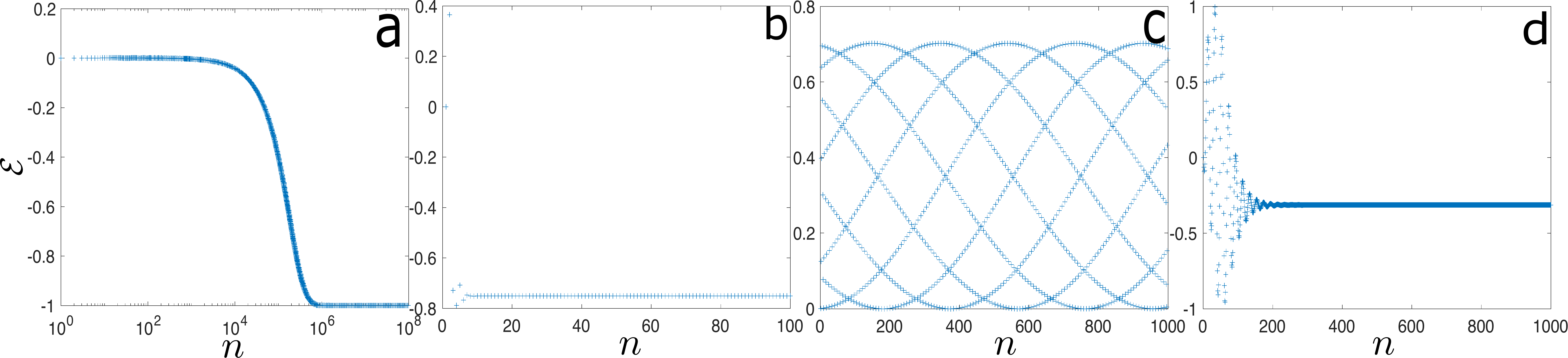}
    \caption{The expected energy of the three-level system versus the measurements step $n$ for $\tau=0.1$(a), $\tau=2$(b), $\tau= 4.31697$(c), and $\tau=4$(d). In subfigure (a), we have $\xi_1 \sim \xi_2 \sim 1$ and $|\xi_1|> |\xi_2|$, the number of steps it takes to reach the state $|\xi_1^R\rangle$ is very large. As a comparison, in subfigure (b), $|\xi_1|\sim 1$ and $|\xi_2| \sim 0.4$, and the transition to state $|\xi_1^R\rangle$ is very fast. In (c), we have $|\xi_1|=|\xi_2|$. Hence the energy oscillates due to the competition between the two states. (d) is close to the oscillation point but not exactly on it, so the amplitude of oscillation decays to zero. }
    \label{fig:three4}
\end{figure*}
The two-level system example, treated in Sec. \ref{sec:two-level}, is obviously a very special case, and misses a lot of the different dynamical regimes. We now consider a three-level system, with no dark states. According to the theory developed, we will have two eigenvalues $\xi_1$ and $\xi_2$, besides $\xi_0=0$, in the unit disk and the long-time limit of the system will be determined by the eigenvalue which is larger in magnitude in absolute value sense. When we vary, say $\tau$, we can find different cases. When $|\xi_1|=|\xi_2|\neq 0$,  the relative phase of these eigenvalues will play a special role, namely we then expect  oscillatory behavior to persist forever. This is investigated here with an example using the charge picture.

We consider a three-level system with the Hamiltonian
\begin{equation}
	H=-\gamma (|0\rangle \langle 1|+ |1\rangle \langle 0| + |1\rangle \langle 2| + |2\rangle \langle 1| + |0\rangle \langle 0| ).
\end{equation}
Note that we have added an onsite energy $-\gamma$ on the node labeled $|0\rangle$.  The spectrum of $H$ is given by $E_i^3 - E_i^2 - 2 E_i +1=0$, where we have set $\gamma=1$. We then find $E_0 \simeq -1.25$, $E_1 \simeq 0.445$, and $ E_2 \simeq 1.80$.  We prepare the system initially in state $|\psi_{\rm in} \rangle = |2 \rangle$ and we perform  null measurements, every $\tau$ units of time, on the detection state $|\psi_{\rm d}\rangle=|0\rangle$. Using Eq. (\ref{charge}), there are three charges: $p_0 \simeq 0.108$, $p_1 \simeq 0.349$, and $p_2 \simeq 0.543$, located at $e^{-i E_0 \tau}$, $e^{-i E_1 \tau}$, and $e^{-i E_2 \tau}$ (see Fig. \ref{fig:threeabs} left panels). This forms our charge picture with two stationary points $\xi_1$ and $\xi_2$ inside the unit disk. Since the energy spectrum is non-degenerate, there are no dark states in the system. As demonstrated in Fig. \ref{fig:threeabs}, varying $\tau$ we get different charge configurations, that yield different behaviors for the measurement process. 

In Fig. \ref{fig:three4}, we show the energy of the system versus $n$ for four choices of $\tau$. Notice the different scales of $n$ in these plots. For small $\tau$, we find Zeno dynamics, the system is lingering in one state for very long time, but eventually switches to a state that is stable in time. The turn over is seen roughly at $n= 10^5$ (see Fig. \ref{fig:three4}(a)).  In contrast, in Fig. \ref{fig:three4}(b), when $\tau=2$, we find a steady state after roughly five measurements. This corresponds to a case where the eigenvalues $\xi_1$ and $\xi_2$ are separated, while in the Zeno case $|\xi_1|\simeq |\xi_2| \simeq 1$. We also see oscillatory behavior for the special choice of $\tau= 4.31697$ in Fig. \ref{fig:three4}(c) or when working close to this value, where the oscillations decays away eventually, Fig. \ref{fig:three4}(d). To gain insights on these behaviors we go back to the charge picture in Fig. \ref{fig:threeabs}. 

 In Fig. \ref{fig:threeabs}, we plot the absolute values of eigenvalues $\xi_1$ and $\xi_2$ and the corresponding charge configurations for $\tau=0.1,\ 2,\ 4.31697,\ 4$. In the Zeno regime,  i.e., $\tau \rightarrow 0$, all the three charges merge (see Fig. \ref{fig:threeabs}(a)). Since the charges are closely situated, it is obvious from electrostatics that the stationary points are all coalescing in the vicinity of the charges but in the unit disk. This is because all the charges are positive.    The eigenvalues of $\hat{\mathfrak{S}}$ are $|\xi_1| \simeq |\xi_2| \simeq 1$. Hence the system reaches the final state slowly, since  clearly both $|\xi_1|^n$ and $|\xi_2|^n$ decay slowly.  This behavior is generic to the Zeno limit and it will be investigated in generality in Sec. \ref{Sec:charge}. A general feature of the eigenvalues is found to be: 
 \begin{equation}
 	|\xi_i| \geq \cos(\Delta E \tau/2),
 	\label{eq:zeno cos}
 \end{equation}
for small  $\tau$, where $\Delta E$ is the difference between the maximum of the energy $E_{max}$ and the minimum $E_{min}$ \cite{Gruenbaum2013}. In our case $\Delta E \simeq 3.05$. We see, with this lower bound, that $\tau \rightarrow 0$ all the eigenvalues of the survival operator approach unity (and hence one cannot simply neglect one compared to the other, unless $n$ is really large). In Fig. \ref{fig:threeabs}(f) we plot this bound for demonstration.

In Fig. \ref{fig:threeabs}(b), i.e., $\tau=2$, we have two charges that are merging \cite{Gruenbaum2013}. This means that the phases satisfy $\exp(-i E_0 \tau) \simeq \exp( -i E_2 \tau)$. From basic electrostatics, if  we have two nearby charges we expect a stationary point in their vicinity, where the forces are balanced. Thus once again we see that the charge picture can be used to rationalise our finding, and more importantly in Sec. \ref{Sec:charge}, we will present a more general theory based on it.  This leads to $|\xi_1| \simeq 1$, while $|\xi_2| \simeq 0.4$. The system reaches the final state very fast, since $|\xi_1| \gg |\xi_2|$. In Fig. \ref{fig:threeabs}(c), $\tau=4.31697$, $|\xi_1|=|\xi_2|$, which leads to the oscillation in Fig. \ref{fig:three4}(c) predicated in Eq. (\ref{eq:quantum dynamics}).

\subsection{Artificial driven atom with $V$-shaped energy structure}\label{sec:V}

\begin{figure}
	\centering
	\includegraphics[width=1\linewidth]{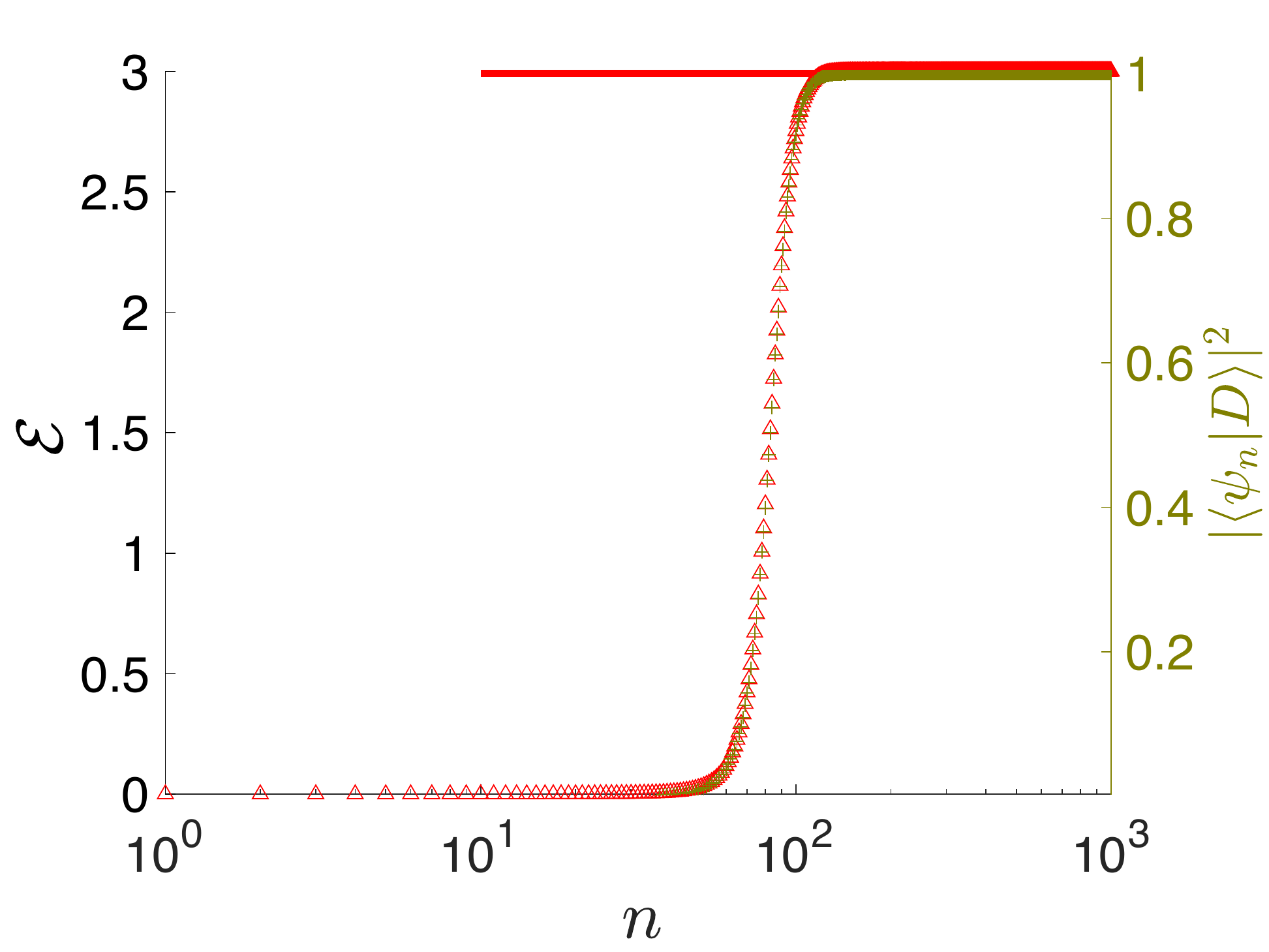}
	\caption{With simulations, we plot the mean energy (red  triangles) of the $V$-shaped system and the probability to be in state $|D\rangle$ (green crosses) versus $n$. Here we choose $\gamma_1 = 0.01$, $\gamma_2=1$, and $\tau = 0.5$. The red line is the theoretical prediction of the mean energy for the large-$n$ limit from Eq. (\ref{eq:38}). The null measurements shelves  the system  in state $|D\rangle$. } 
	\label{fig:three-energy}
\end{figure}
We now consider an artificial atom system with the famous $V$-shaped energy level structure that allows for shelving \cite{Plenio1998}, see Fig. \ref{fig:three}. 
The system's states are $|G \rangle$ (ground state) and $|D\rangle$ and $|B\rangle$. Inspired by the experiment in Ref. \cite{Minev2019}, we will investigate  the case where transition amplitudes from $|G\rangle$ to $|D\rangle$ and $|G\rangle$ to $|B\rangle$ vary considerably.  The system starts in the ground state, and then conditional null measurements are made in state $|B\rangle$. The question is where will we find the particle, when $n$ is large? It turns out that the particle is shelved in state $|D\rangle$ and the amplitude of finding the particle in the ground state diminishes with $n$. In some sense the fact that we gain knowledge from measurements that the system is not in $|B\rangle$ implies that it cannot be found in $|G\rangle$ also, since $|G\rangle$ is loosely speaking the doorway to $|B\rangle$.  Thus the energy of the system is going to increase from the energy of the ground state to the energy of state $|B\rangle$. We now investigate this scenario in more detail. The Hamiltonian $H$ reads:  
\begin{align}
     H &=E_D |D\rangle \langle D|+E_B |B\rangle \langle B| +E_G |G\rangle \langle G| \nonumber\\
    &+\gamma_1(|G\rangle \langle D|+|D\rangle \langle G|)+\gamma_2( |G\rangle \langle B|+ |B\rangle \langle G|).
\end{align}
We first find the eigen energies and vectors of $H$ using the basis  $\{|D\rangle, |G\rangle, |B\rangle\}$. To simplify the calculations, we set $E_G=0, E_D=3, E_B=5, \gamma_2 =1$. We keep $\gamma_1$ as a free parameter, and we will consider the limit when it approaches zero.  Then the eigenvalues $E_i$ of $H$ are given by $E_i^3 - 8 E_i^2 + (14 - \gamma_1^2) E_i + 3 +5 \gamma_1^2 = 0$. When $\gamma_1 \rightarrow 0$, we have $(-3 + E_i) (-1 - 5 E_i + E_i^2) \simeq 0$, which leads to $E_0 \simeq -0.2$, $E_1 \simeq 3$, and $E_2 \simeq 5.2$. The eigenstates of $H$ are $|E_i\rangle = N\{ (-1 - 5 E_i + E_i^2)/\gamma_1, E_i -5, 1 \}^T$, where $N$ is the normalization. As $\gamma_1 \rightarrow 0$, $|E_1\rangle \simeq |D\rangle$, a result we will use later. 

We then choose $\gamma_1 =0.01$. By the definition of charges, we have $p_0 = |\langle B|E_0\rangle|^2 \simeq 0.03576  $, $p_1 = |\langle B|E_1\rangle|^2 \simeq  2.040 \times 10^{-6} $, $p_2= |\langle B|E_2\rangle|^2 \simeq  0.9642 $. So $p_2 \gg p_0 \gg p_1$, the charges $p_0$ and $p_1$ are much weaker than $p_2$.  In the electrostatic language, we have two weak charges, and as a result, the stationary points $\xi_1$ and $\xi_2$ will approach these weak charges separately. With simple numerical calculations, we find the eigenvalues of $\hat{\mathfrak{S}}$ are $\xi_1 \simeq 0.0707+0.9975i$, $\xi_2=0.9292+0.0742i$, so $|\xi_1| \simeq |\xi_2| \simeq 1$, and $|\xi_1|\textgreater |\xi_2|$.

We see that we have a very weak charge $p_1$ in the system. In this case we expect and indeed find an eigenvalue of the survival operator that is very close the vicinity of this charge, but in the unit disk, and it is $\xi_1$. This can be understood using basic electrostatics, namely for three positive charges on the unit disk, we expect to find a stationary point close to the weak charge, the remaining charges balance the force (Similarly, the stationary point between Earth and Moon  is closer to the Moon).  We will show this in more generality in Sec. \ref{Sec:charge}, using perturbation theory, but for now we point out that to leading order we expect that the corresponding eigenstate, namely $|E_1\rangle$, is selected as the stationary state of the system.  So for a quantum particle initially in the ground state $|G\rangle$,  using Eq. (\ref{eq:weakfinal}), we have for large $n$
\begin{equation}
	|\psi_f\rangle \simeq e^{i n \phi}|\xi_1\rangle \simeq  e^{-i n E_1 \tau} |E_1\rangle \simeq e^{-i n E_1 \tau} |D \rangle.
	\label{eq:38}
\end{equation}
Secondly,  because we have two weak charges, i.e. both $p_0$ and $p_1$ are weak compared to $p_2$, that leads to $|\xi_1| \simeq |\xi_2|$ as mentioned. Therefore, for not too large $n$, and from Eq. (\ref{final}), both the eigenvalues contribute, and the transition to the final state Eq. (\ref{eq:38}) is slow.  Only at a certain critical large number of measurements do we observe a transition.

 As shown in Fig. \ref{fig:three-energy}, the expected energy of the monitored system exhibits a transition from the zero energy of ground state $|G\rangle$ to the energy of state $|D\rangle$, which is 3. This transition is found at $n\simeq 50$, which is related to the relative magnitude of $|\xi_1|$ and $|\xi_2|$, since in this example both are close to unity as mentioned.  We also plot the probability of being in state $|D\rangle$ which exhibits a similar transition in the expected energy from zero to unity.
 
As already mentioned in the introduction, we have been inspired by the experiment of Minev \textit{et al.} \cite{Minev2019}. There the steering by conditional measurements was used to control and reverse the quantum jump in mid-flight. This, in turn, is based on concepts of quantum trajectories, well investigated in quantum optics. There the emission events, or more generally quantum jumps, are an inherit to the dynamics of the system, for example, a fluorescence process. In our approach, which is different from the experiment, we impose the jumps and the conditional steering by repeated projective measurements, where the condition is a null measurement (i.e., non-detection).  We believe that due to the advances in single-particle manipulation, the proposed method will be useful both in the steering of single-particle systems, as shown here, but also for their control.  For example, in the quantum search problem, the goal of the field is to speed up the search, which in turn is presented as the time it takes for a quantum walker to reach the target state. Then if we detect many events of failed measurements (like those analyzed here), we may wish to restore to control. By that, we mean that after observing many failed detections, the system in the search process is pushed toward a dark state. These dark states, in systems with built in symmetry, are stable. This means that after the target is not detected for a while, it will be of benefit to add control, for instance, to restart the search process or add an external perturbation. This approach and its effect on the quantum search process must be investigated in more detail in the future. 
 
 {\bf Remark:} The state $|D \rangle$  in this section is traditionally called dark, indeed this term is obviously very physical. In the context of this work, and based on our definition of the dark state $|\delta \rangle$, note that state $|D\rangle$ is only nearly dark. We defined  dark states, as such  that starting in those states the system is never detected, while here a system initially starting in the state $|D \rangle$ can be detected, at least in principle (without conditioning the measurements, of course). The meaning of nearly dark is related to the  set of parameters like $\gamma_1$ we and others use in these problems. The meaning of a nearly dark state is explored below in perturbation  theory in full generality using the so called weak charge theory (see Sec. \ref{sec:weak}).

\section{Glued binary Tree}\label{sec:tree}
\begin{figure}
    \centering
    \includegraphics[width=0.99\linewidth]{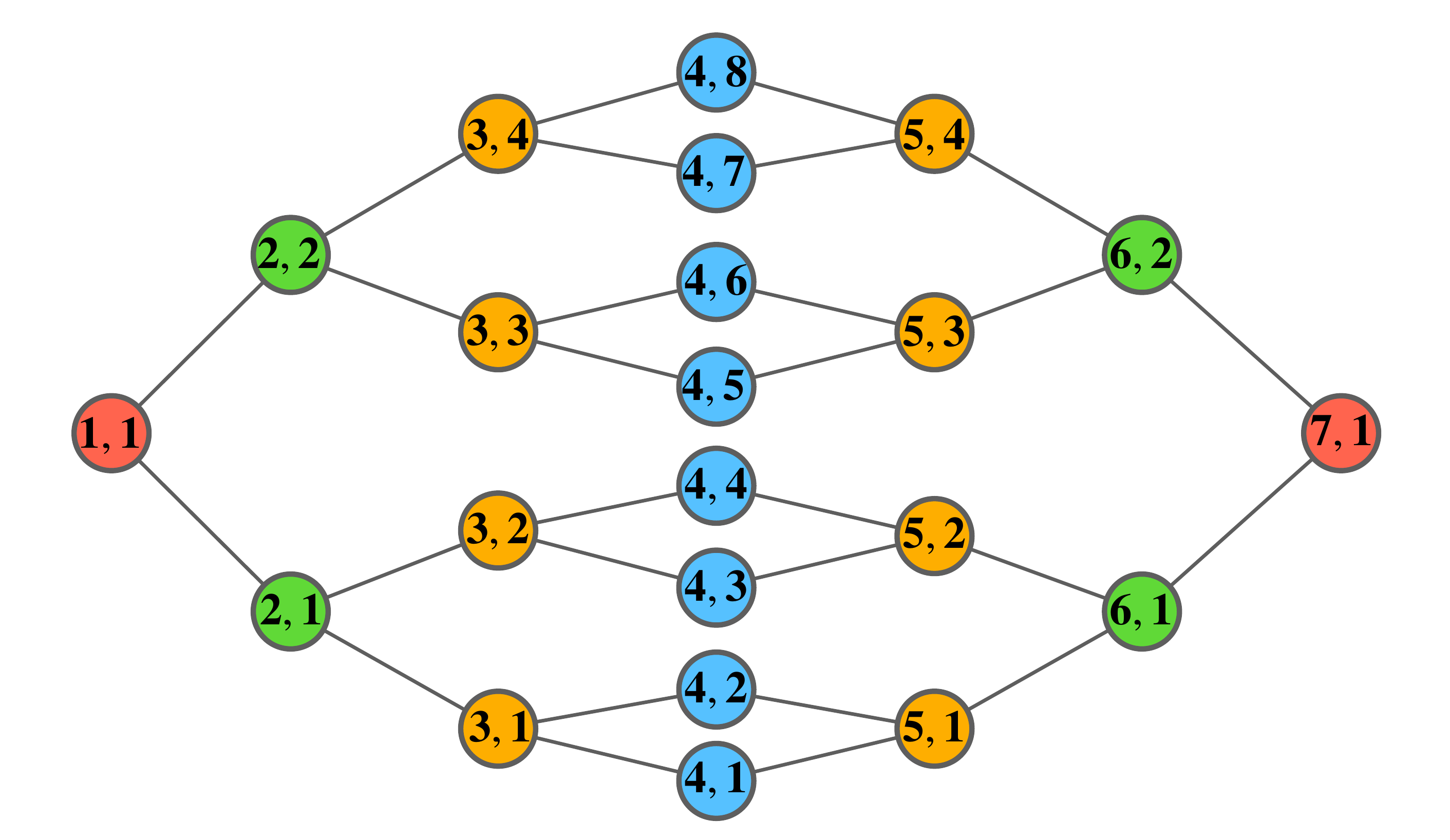}
    \caption{Glued binary tree graph $G_3$. The detector is set on the state $|1,1\rangle$. We consider different initial states. }
    \label{fig:G4tree}
\end{figure}

\begin{figure*}
    \centering
    \includegraphics[width=0.7\linewidth]{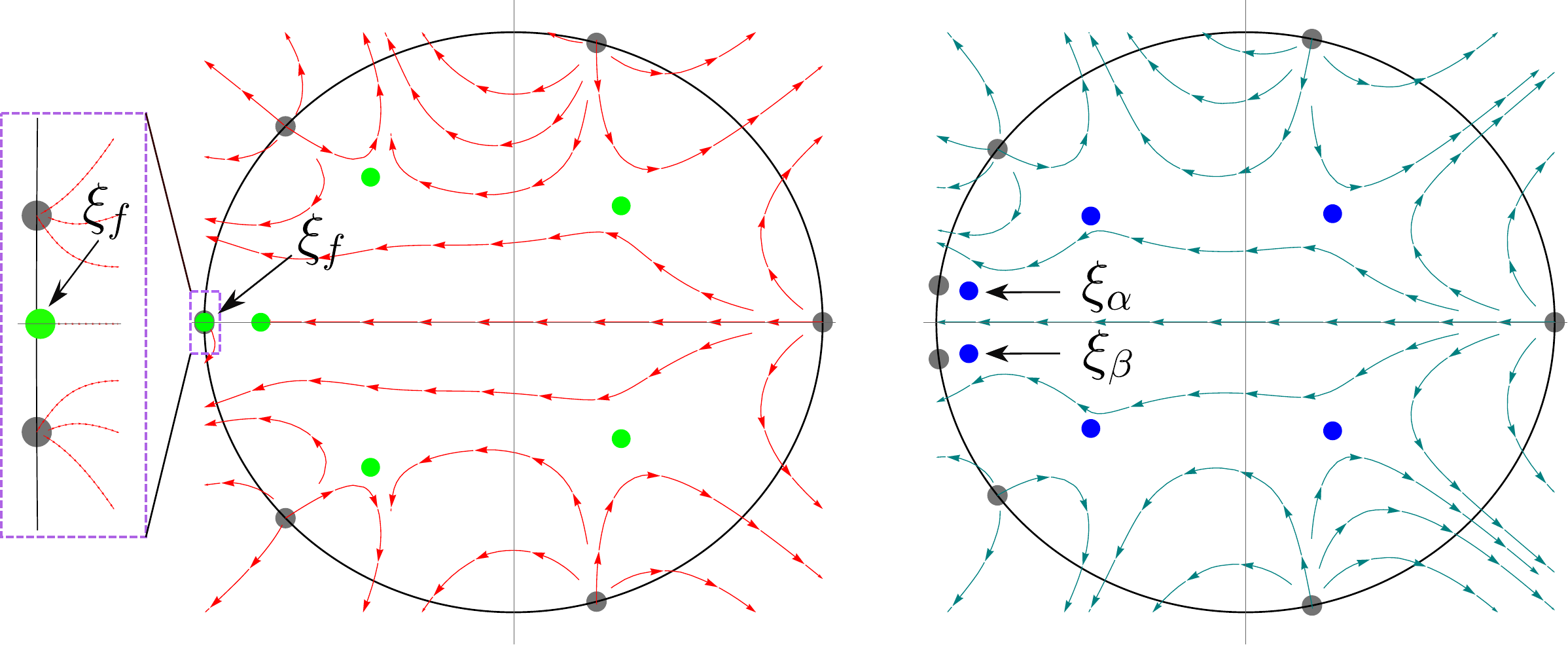}
    \caption{ The classical charge pictures for the eigenvalues of $\hat{\mathfrak{S}}$ of $G_3$ tree model when $\tau=1.2$ (left) and $\tau=1.25$ (right). As shown in figure, when $\tau=1.2$, there is only one eigenvalue $\xi_f$ that is closest to the unit circle. While at $\tau=1.25$, two eigenvlaues $\xi_{\alpha}$ and $\xi_{\beta}$ that are equally close the the unit circle are found. In the latter case we anticipate quantum dynamics in the long-time limit, while in the former a unique steady state emerges. In this system we have 11 distinct  energy levels, see Table \ref{tab:1}. One would naively expect to find 11 charges, however  some charges vanish, namely the detected state $|1,1\rangle$ is orthogonal to some of the energy levels, due to symmetry. As a result we have 7 charges.  }
    \label{fig:G3charge}
\end{figure*}

\begin{figure}
    \centering
    \includegraphics[width=1\linewidth]{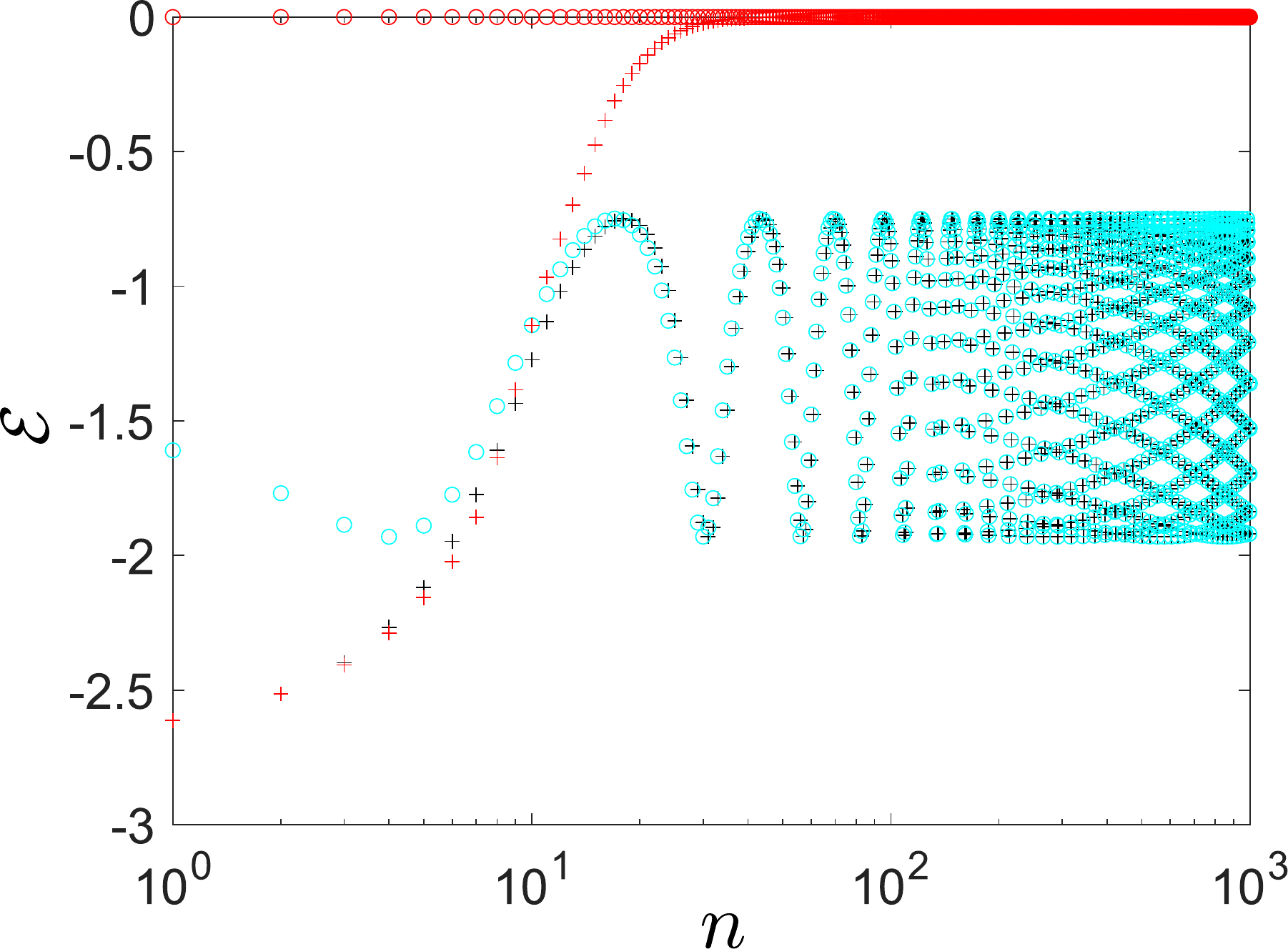}
    \caption{The expected energy of the $G_3$ tree system versus the measurement steps $n$ conditioned the detections are null.  We prepare the system in its ground state ($\langle {\cal E} \rangle = -2.61313$),  and then perform repeated stroboscopic measurements at the state $|\psi_{\rm d}\rangle=|1,1\rangle$ (see Fig. \ref{fig:G4tree}). As demonstrated in Fig. \ref{fig:G3charge}  when $\tau=1.2$, there is only a unique eigenvalue $\xi_f$ of $\hat{\mathfrak{S}}$  that is closest to the unit circle, corresponding to the state $|\xi_f^R\rangle$, with expected energy $\langle {\cal E}_f\rangle=0$ (red circles). As shown here, the energy of the system (red crosses, numerical simulations) reaches the expected energy when $n \simeq 20$. The detections pump the energy of the system and send it to the new steady state. When $\tau=1.25$, there are two eigenvalues $\xi_{\alpha}$ and $\xi_{\beta}$ that are equal in magnitude $|\xi_b|=|\xi_a|$ (see Fig. \ref{fig:G3charge}), while they are also the largest eigenvalues (besides the darks states which are not relevant due to the initial condition). Now the total energy of the system oscillates. Our theory (cyan circles) predicts this oscillation and perfectly matches the numerical simulations (black crosses) when $n$ is large. }
    \label{fig:G3energy}
\end{figure}

\begin{table*}
	\begin{tabular}{l|ccccccccccc}
\hline
& $E_0$ & $\ E_1\ $ & $\ E_2\ $ & $\ E_3$ & $E_4$ & $\ E_5\ $ & $\ E_6\ $ & $\ E_7\ $ & $\ E_8\ $ & $\ E_9\ $ & $\ E_{10}\ $ \\
\hline
Energy  & $-2\sqrt{2}\cos(\pi/8)$ & $-\sqrt{6}$ & $-2$ & $-\sqrt{2}$ & $-2\sqrt{2}\sin (\pi/8)$ & $0$ &  $-E_4$ & $-E_3$ & $ -E_2$ & $-E_1$ & $-E_0$ \\
Degeneracy & 1 & 1 & 3 & 1 & 1 & 8 & 1 & 1 & 3 & 1 & 1 \\
$\langle \psi_{\rm d}|E_{k ,i}\rangle=0$ & 0 & 1 & 0 & 1 & 0 & 5 & 0 & 1 & 0 & 1 & 0  \\
$\langle \psi_{\rm d}|E_{k ,i}\rangle \neq 0$ & 1 & 0 & 3 & 0 & 1 & 3 & 1 & 0 & 3 & 0 & 1  \\
\hline
\end{tabular}
\caption{The energy spectrum of $G_3$ tree. The detection state $|\psi_{\rm d}\rangle =|1,1\rangle$. The number of states with zero and non-zero overlap with the detection state $|\psi_{\rm d}\rangle$ is listed in the last two lines. }
\label{tab:1}
\end{table*}

As another application of our general theory, we consider the glued binary tree \cite{GluedTree1,GluedTree2}. Glued trees were investigated previously as they provide exponential speedup for quantum search algorithms \cite{Farhi1998,10.1145/780542.780552}, and this was observed in a recent experiment \cite{Shi:20}.  In contrast we consider the effect of null measurements on this popular model. 

\subsection{Eigenstates and eigenenergies}

First, let us define a sequence of graphs $G_d$. $G_d$ consists of two balanced binary trees \cite{GluedTree1}. The total number of vertices in $G_d$ is $2^{d+1}+2^d-2$. In Fig. \ref{fig:G4tree} we present the $G_3$ tree, which will serve as our example. This graph describes the Hamiltonian of the system, namely the nodes are the states and the links represent the hoping amplitudes between states. All transitions are identical, so that the Hamiltonian is  given by the adjacency matrix of the graph. 

 The eigenenergies  and eigenstates of this model  were obtained in Refs \cite{GluedTree1,GluedTree2}. Here we briefly recap the solution of the stationary  Schr\"{o}dinger equation. We consider a labeling along the ``columns" and ``rows" of the form $(j,s)$ (see Fig. \ref{fig:G4tree}), where $j$ goes from  $0$ to $2d$ indicates the distance from the left root to right root along the graph, and $s$ goes from $0$ to $N_{d,j}-1$ is the location within a given column. $N_{d,j}$ is the number of sites in a given column $j$, and it is given by $N_{d,j}=2^j$ for $j\leq d$, and $N_{d,j}=2^{2d-j}$ for $j\textgreater d$. Then the eigenstates of $H$ are \cite{GluedTree2}:
\begin{equation}
	|E_{k,d-v, \alpha}\rangle = \frac{1}{\sqrt{d-v+1}}\sum_{j=0}^{2(d-v)} \sin[\frac{k(j+1)\pi}{2(d-v+1)} ]|\text{scol} \ j; \alpha ,v\rangle
	\label{Get}
\end{equation}
with
\begin{equation*}
	|\text{scol} \ j; \alpha ,v\rangle = \sum_{s=2\alpha N_{j,d-v}}^{(2\alpha+1)N_{j,d-v}-1} \frac{|j+v,s\rangle - |j+v, s+N_{j,d-v}\rangle}{\sqrt{2N_{j,d-v}}}
\end{equation*}
 in which $|j,s\rangle$ denote the Hilbert space vector associated with vertex $(j,s)$;
and $v=0, \cdots, d$; and $k=1, \cdots, 2(d-v)+1$. In $|\text{scol} \  j; \alpha ,v\rangle$, $j=0, \cdots, 2(d-v)$ and $\alpha=0, \cdots, 2^{v-1}-1$. $\alpha$ is an integer, so when $d=0$, $\alpha=0$. The corresponding eigenvalues are given by   \cite{GluedTree2}
\begin{equation}
	E_{k,d-v}=-2\sqrt{2} \cos(\frac{k \pi}{2(d-v+1)} ).
	\label{Gev}
\end{equation}

\subsection{Null measurements}

Due to the symmetry of the glued binary tree, the spectrum exhibits a large degeneracy. From Eq. (\ref{darkstateeq}), there are many dark states in the system. For example, the energy level $E = 0$ has a multiplicity of $2^d$, and hence with Eq. (\ref{darkstateeq}),  we can construct $2^d-1$ dark states in this energy subspace $E=0$. If the initial state of the quantum particle has some overlap with the dark subspace, it can be trapped in the dark subspace. In this case the dynamics will be determined by the degenerate eigenenergies as mentioned.

To present the quantum dynamics via measurements, we treat a specific model. We choose the $G_3$ tree as shown in Fig. \ref{fig:G4tree}, which consists of $22$ nodes.  So $d=3$. We use Eq. (\ref{Gev}) and for simplicity we label the energy levels according to $ E_0 \textless E_1 \textless E_2 \cdots \textless E_{10}$, see Table \ref{tab:1}. The energy level $E_{5}=0$ is 8-fold degenerate and $E_2=2$ and $E_7=-2$ are 3-fold degenerate. Other states are not degenerate. The detection state is $|\psi_{\rm d}\rangle=|1,1\rangle $ (see Fig. \ref{fig:G4tree}), the eigenstates $|E_1\rangle$, $|E_3\rangle$, $|E_7\rangle$, $|E_9\rangle$ and $|E_{5,i}\rangle$, where $ i=4,5,6,7,8$, have no overlaps with the detected state $|\psi_{\rm d}\rangle$, they are dark states by definition. Then we are left with 3 energy levels that are 3-fold degenerate, i.e., $E_2=2$, $E_7=-2$, and $E_{5}=0$. Using Eq. (\ref{darkstateeq}), we construct from each degenerate energy subspace two dark states. For instance, the dark states on energy level $E_2$ are $|\delta_{2,1}\rangle =\sqrt{1/6}|E_{2,1}\rangle- \sqrt{5/6} |E_{2,2}\rangle  $ and $|\delta_{2,2}\rangle =1/\sqrt{6}|E_{2,1}\rangle+1/\sqrt{30}|E_{2,2}\rangle-2/\sqrt{5}|E_{2,3}\rangle $. Following the same procedure, we can also construct the dark states $|\delta_{5,1}\rangle$, $|\delta_{5,2}\rangle$, $|\delta_{8,1}\rangle$, and $|\delta_{8,2}\rangle$ using Eq. (\ref{darkstateeq}). Then we have all 15 dark states in the system.

\subsubsection{Initial condition orthogonal to the dark states}

Now, based on the charge picture approach, we show different quantum dynamics driven by repeated measurements.  We start from the ground state of the system, namely, $|\psi_{\rm in}\rangle=|E_0\rangle$. The ground state is an eigenstate of Hamiltonian $H$ and non-degenerate, hence the initial state has no overlap with all the dark states, i.e., $\langle \psi_{\rm in}|\delta_{k,j}\rangle = \langle E_0 |\delta_{k,j}\rangle=0$. This is obvious since the energy eigenstates are orthogonal with respect one to another. For the first case, we consider $\tau=1.2$. The eigenvalues $\xi_i$ of survival operator $\hat{\mathfrak{S}}$ are stationary points of the charge field in Fig. \ref{fig:G3charge}. There is a unique maximum $\xi_f \simeq -0.999767$ of the set $\{|\xi_i|\}$ that is closest to the unit circle (see Fig. \ref{fig:G3charge}). So the system will approach this fixed point when the detection number is large. From Eq. (\ref{eq:bright only}), the final state is $|\psi_f\rangle = e^{i n \phi_f} |\xi_f^R\rangle$, whose expected energy is ${\cal E}=\langle \xi_f^R|H|\xi_f^R\rangle= 0$, while initially the energy is ${\cal E}_0 \simeq -2.61313$. The measurements transfer energy to the quantum system continuously until the system reaches its equilibrium state. We present this in Fig. \ref{fig:G3energy}.

For the second scenario, we choose $\tau =1.25$, where there are two eigenvalues $\xi_{\alpha} \simeq -0.894962 + 0.108282 i$ and $\xi_{\beta} \simeq  -0.894962 - 0.108282 i $ that are equally close to the unit circle (see Fig. \ref{fig:G3charge}). The expected energy of the corresponding eigenstates are $ {\cal  E}_{\alpha}  \simeq 1.46103 $ and  $ {\cal E}_{\beta}  \simeq -1.46103 $. Since there are two eigenvalues equally close to unit circle, the system will oscillate in time. Due to the competition between the states $|\xi_{\alpha}^R\rangle$ and $|\xi_{\beta}^R\rangle$ during the measurements process the energy oscillates in the detection process (see Fig. \ref{fig:G3energy}).  The measurement transfers energy to the quantum system and back in an oscillatory  way, and the system never reaches a stationary state. These interesting phenomena reveal the effect of the measurements and are also instructive for the quantum dynamics out of equilibrium.

\subsubsection{Initial condition  overlapping with the dark states}
\begin{figure}
    \centering
    \includegraphics[width=1\linewidth]{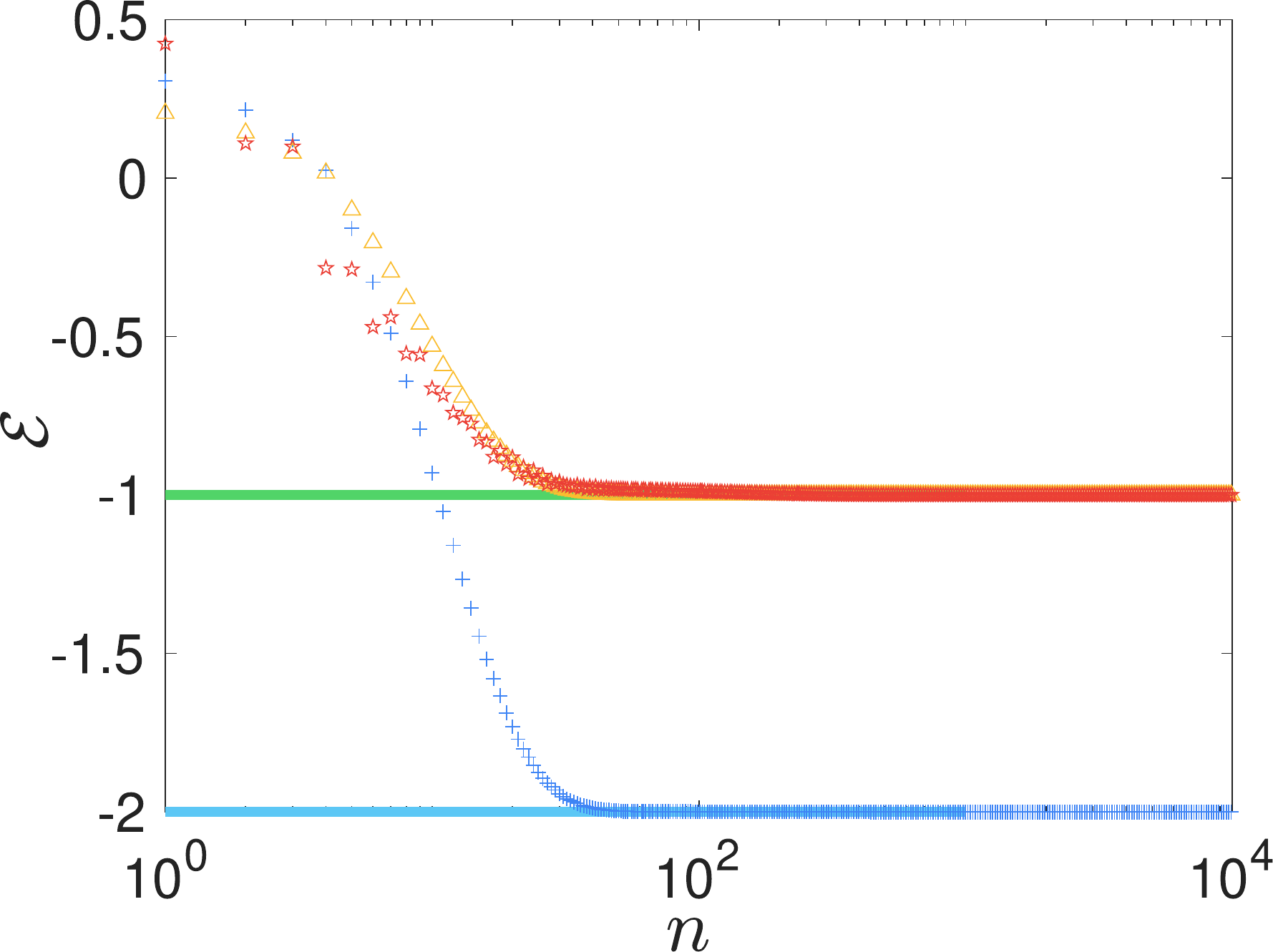}
    \caption{The expected energy of the $G_3$ tree system versus the measurement steps $n$. The blue crosses are numerical simulations for the initial state $|\psi_{\rm in}\rangle=(|E_{2,1}\rangle+|E_{10}\rangle)/\sqrt{2}$. The selection rule for the final energy of the system  indicates, that this final energy is the same as that of state $|E_{2,1}\rangle$ since that state belongs to a degenerate subspace, while $|E_{10}\rangle$  does not.  And the theoretical value of ${\cal E}$ in the large-$n$ limit is the blue line. The yellow triangles are for the initial state $|\psi_{\rm in}\rangle=(|E_{2,1}\rangle+|E_{10}\rangle+|E_{5,1}\rangle )/\sqrt{3}$ (here both $|E_{2,1}\rangle$ and $|E_{5,1}\rangle$ are degenerate, hence the final state is different from the previous case). We also show the mean energy for  $|\psi_{\rm in}\rangle=(|E_{2,1}\rangle+|E_{10}\rangle+|E_{5,1}\rangle +|E_6\rangle )/2$ in red pentagrams. The theoretical predictions of the mean energy ${\cal E}$ is the green line.  Note that red and yellow curves merge in the large-$n$ limit. This is because the difference between these two initial conditions, is an energy eigenstate which is non degenerate, which is unimportant in the long-time limit.}
    \label{fig:G3dark}
\end{figure}

Now we consider the cases that the initial state has some overlap with the dark states. According to Eq. (\ref{eq:dark only}), the final state of the system is determined by the dark states, which are constructed from the degenerate energy levels. In this case we expect that the non-degenerate eigenenergies will be irrelevant in the long-time limit.  In this part, we exploit the degeneracy of $H$ and explore the selection rule for the degenerate states.

We first consider the case that the initial state $|\psi_{\rm in}\rangle=(|E_{2,1}\rangle+|E_{10}\rangle)/\sqrt{2}$, which is a linear combination of the degenerate energy level $E_2$ and non-degenerate energy level $E_{10}$. The mean energy of the $G_3$ tree system at the beginning is ${\cal  E}_0=E_2/2+E_{10}/2 \simeq 0.31$. Conditioning the measurements to be null, the system converges to the degenerate energy level only and the component of the non-degenerate energy level is eventually wiped out. So, the expected energy of system in the final state is ${\cal E}= E_2=-2$. In Fig. \ref{fig:G3dark}, our numerical simulations show that the energy of the system is indeed $E_2$ when $n$ is large.

 Now we add the component of another degenerate energy level $E_5$  to the initial state. The initial state then reads $|\psi_{\rm in}\rangle=(|E_{2,1}\rangle+|E_{10}\rangle+|E_{5,1}\rangle )/\sqrt{3}$. This is a linear combination of two degenerate energy levels and one non-degenerate energy level. Following the selection rule, the two degenerate energy levels determine the final state of the system, and the mean energy of the system will reach the value ${\cal E}=-1$ eventually. If we add another non-degenerate energy level $E_6$, i.e., $|\psi_{\rm in}\rangle=(|E_{2,1}\rangle+|E_{10}\rangle+|E_{5,1}\rangle +|E_6\rangle )/2$, this does not change the final state, as the selection rule  of the measurements wipes out both the $|E_{10}\rangle$ and $|E_6\rangle$ components. We present the numerical simulations in Fig. \ref{fig:G3dark}, and the results are consistent with our predictions.

\section{Null Measurements, insights from the charge theory}\label{Sec:charge}
We now provide general insights from the charge theory, which are used to find the largest eigenvalue of $\hat{\mathfrak{S}}$ in generic situations. The technique here presented follows, and in some cases extends the ideas in \cite{Gruenbaum2013,Yin2019,PhysRevResearch.2.033113}, that were developed in the context of the first detection problem.  We consider cases where the dark space is empty, simply because a system with a non-empty dark space can be treated exactly with the tools given in Eq. (\ref{darkstateeq}). So, in this section $|\xi_i| \textless 1$. Furthermore, we consider cases where one or several eigenvalues of the survival operator are close to the unit circle, and hence these are the largest. We develop an approximate expression for the eigenvalue(s), and also give insight on the eigenvectors. We consider four cases: i) a weak charge scenario, ii)  two charges merging picture, iii) triple-charge theory, and finally iv) quantum Zeno regime, where all the $\xi$s approach the unit circle.

\subsection{Weak charge theory}\label{sec:weak}
Assume that one of the overlaps denoted $p_0$, associated with energy level $E_0$, is small, $p_0 \ll 1$, and in particular much smaller than all the others. In the electrostatic language, we have a weak charge at $\exp(-i E_0 \tau)$. We find a stationary point close to this charge, denoted $\xi_f \simeq \exp(-i E_0 \tau)$. At $\xi_f$, the electrostatic force vanishes, because the force of the weak charge balances all other forces.  By analogy, the stationary point of the Moon-Earth system is much closer to the Moon than to the Earth. Using Eq. (\ref{charge}) together with perturbation theory, we get \cite{Yin2019}
 \begin{equation}
	 \xi_f \sim e^{-i E_0 \tau}-\epsilon,
	 \label{eq:46}
\end{equation}
where
\begin{equation}
\epsilon \sim \frac{p_0}{\sum_{k\neq 0} p_k/(e^{-i\uptau E_0}-e^{-i\tau E_k})}.
\end{equation}
Since $p_0 \ll 1$, $\epsilon \sim 0$. The leading term of $\xi_f$ is $e^{-i E_0 \tau}$, hence $|\xi_f| \sim 1$ and $|\xi_f |<1$. From Eq. (\ref{eq:bright only}), for a system with such a weak charge,    the final state is  $|\psi_f\rangle \sim e^{i n \phi_f} |\xi_f^R\rangle $.  Substituting $\xi_f$ into Eq. (\ref{righte}), the expression for the right eigenstate $|\xi_f^R\rangle$ can be highly simplified, which leads to 
\begin{equation}
	|\psi_f \rangle \sim  e^{i n \phi_f}   |E_{0}\rangle,
	\label{eq:weakfinal}
\end{equation}
where $\phi_f = -i \ln(\xi_f/|\xi_f|)\sim -E_0 \tau $. Eq. (\ref{eq:weakfinal}) indicates that the final state is the energy eigenstate $|E_0 \rangle$. And the global phase accumulated is approximately the energy $E_0$ multiplied by the evolution time. The repeated measurements drive the system to this specific state.  To get a deeper understanding of this result, we go back to the definition of the weak charge $p_0$, where $p_0 = |\langle \psi_{\rm d} |E_0\rangle|^2 \ll 1$. Actually, the ``weakness'' of the charge means the overlap of the energy state $|E_0\rangle$ and the detection state $|\psi_{\rm d}\rangle$ is nearly zero, i.e., $\langle \psi_{\rm d}| E_0\rangle \sim 0$. Hence the bright state $|E_0\rangle$ is acting like a nearly dark sate, due to the small overlap. The expected energy of the final state is $ \langle {\cal E} \rangle= \langle \psi_f |H | \psi_f \rangle \sim E_0$.

This picture, was demonstrated already in specific $V$-shaped system in Sec. \ref{sec:V}. More specifically, in this example, the weak charge $p_0 \simeq 2.040 \times 10^{-6}$ (note in Sec. \ref{sec:V}, we denoted it $p_1$). Using Eq. (\ref{eq:46}), we have $\epsilon \simeq 1.945 \times 10^{-6} -1.077 \times 10^{-6} i $,  from here $\xi_1 \simeq 0.0707353 -0.997496 i$ in excellent agreement with the exact value, $\xi_1 =0.0707353 - 0.997494 i$. Moreover, the general Eq. (\ref{eq:weakfinal}) is directly  demonstrated with Eq. (\ref{eq:38}).

\subsection{Two merging charges}
Another mechanism leading to the eigenvalue of $\hat{\mathfrak{S}}$ being close to the unit circle is the case when two energy levels, denoted $E_a$ and $E_b$, satisfy the resonance condition $\exp{(-i E_a\tau)}\simeq \exp{(-i E_b \tau)}$ \cite{Gruenbaum2013}. Note that this can be achieved by modifying $\tau$ or some other parameter of $H$.  We then have two charges $p_a$ and $p_b$ close to each other, both located on the unit circle. So we expect to find a stationary point, denoted $\xi_f$ in their neighborhood. This is because the point of zero force is largely determined by this pair. In analogy, the equilibrium point between two neighboring stars is determined to leading order by these and not by other distant stars. An example was shown in Fig. \ref{fig:threeabs}(b), however now we treat the problem in generality. We need to find an approximation for $\xi_f$ as $\delta \rightarrow 0$, where $\delta = (E_b \tau-E_a\tau )/2\ \text{mod}\ 2 \pi$, which measures the angular distance between the two phases. Using Eq. (\ref{charge}), we find \cite{PhysRevResearch.2.033113}:
\begin{equation}
    \xi_f \sim \frac{p_a e^{-i E_b \tau} +p_b e^{-i E_a \tau}}{p_a + p_b} +O(\delta^2).
    \label{eq:chargetwo}
\end{equation}
As shown in Eq. (\ref{eq:chargetwo}), the charges $p_a$ and $p_b$ determine the location of the stationary point $\xi_p$, and the other charges give only a second order perturbation. Since $\exp{(-i E_a\tau)}\simeq \exp{(-i E_b \tau)}$, $| \xi_f | \sim 1$.  As a demonstration of Eq. (\ref{eq:chargetwo}), we consider the charge configuration in Fig. \ref{fig:threeabs}(b). Using Eq. (\ref{eq:chargetwo}), we have $\xi_f \simeq -0.8170 + 0.5726 i$. While the exact value is $-0.8158 + 0.5722 i$. Here $\delta = 0.09$, and the error comes from terms which are second order in $\delta$.

From Eq. (\ref{eq:bright only}), the final state of the system then is $|\psi_f \rangle \sim e^{i n \phi_f} |\xi_f^R\rangle$.  Substituting $\xi_f$ into Eq. (\ref{righte}), we have
\begin{equation}
    |\xi_f ^R \rangle \sim N ( \frac{\langle E_a|\psi_{\rm d}\rangle }{p_a}|E_a\rangle -  \frac{\langle E_b|\psi_{\rm d}\rangle }{p_b}|E_b\rangle),
\end{equation}
where $N$ is for normalization\footnote{When $\langle E_a|\psi_{\rm d}\rangle$ is real we have  of course $\langle E_a|\psi_{\rm d}\rangle /p_a = 1/\sqrt{p_a}$. 
}. The eigenstate $|\xi_f ^R \rangle$ is a linear combination of the energy eigenstates $|E_a\rangle$ and $|E_b\rangle$ that are merging. As a reminder, when the energy level $E_0$ is 2-fold degenerate, the dark state is given by  $|\delta_{0,1}\rangle = N(\langle \psi_{\rm d} |E_{0,2}\rangle |E_{0,1}\rangle - \langle \psi_{\rm d} |E_{0,1}\rangle |E_{0,2}\rangle)$. Since the energy levels $E_a$ and $E_b$ satisfy the resonance condition $\exp{(i E_a\tau)}\simeq \exp{(i E_b \tau)}$, they now act like one degenerate energy level. However, now they are weighted by the magnitude of the corresponding charges for each energy level.  The expected energy of the final state is
\begin{equation}
	\langle {\cal E} \rangle  \sim \frac{p_b E_a + p_a E_b}{p_a + p_b}.
	\label{eq:51}
\end{equation}
Following the example in Fig. \ref{fig:threeabs}(b), we have $\tau=2$ and the merging charges are $p_0 \simeq 0.108$ and $p_2 \simeq 0.543 $ corresponding to the energy levels $E_0 \simeq -1.25$ and $E_2 \simeq 1.80$. Using Eq. (\ref{eq:51}), the mean energy of the system in the large-$n$ limit is ${\cal E} \simeq -0.744$. The exact mean energy is $-0.7511$, which is presented in Fig. \ref{fig:three4}(b).

The global phase accumulated in the measurements is:
\begin{equation}
	\phi_f \sim -\frac{E_a+E_b}{2}\tau + \frac{p_b -p_a}{p_a+p_b} \delta.
\end{equation}
The first term on the lhs is the leading part of $\phi_f$. Like the final state, the phase is also determined by the two merging energy levels.

\subsection{Triple-charge theory}

\begin{figure*}
    \centering
    \includegraphics[width=0.7\linewidth]{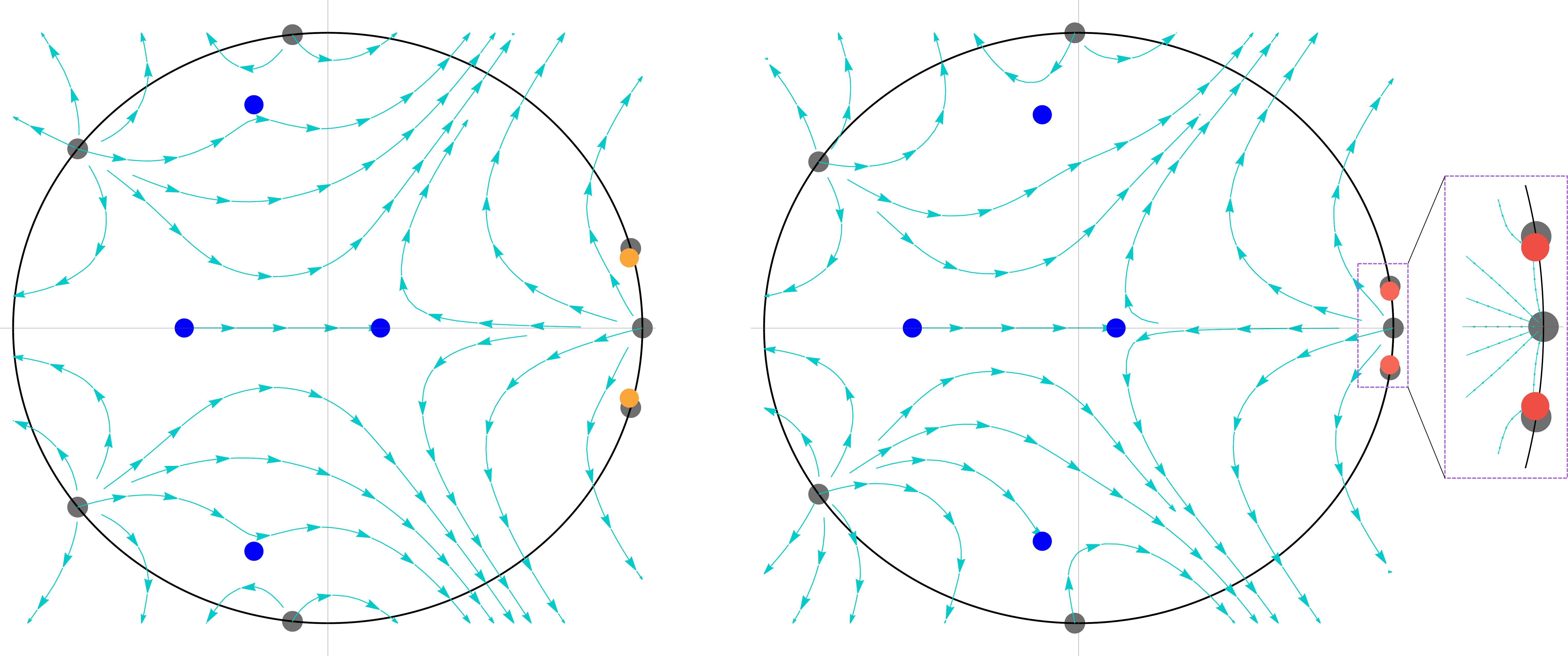}
    \caption{The classical charge pictures for the eigenvalues of $\hat{\mathfrak{S}}$ of $G_3$ tree model when $\tau=2.3$ (left) and $\tau=2.35$ (right). Here $|\psi_{\rm d}\rangle=|1,1\rangle$.  As shown in figure, on the east part, on and just above and below the equator, we see three charges that are close one to another (charges are in gray, on the unit circle).  As a result, there are two eigenvalues $\xi_+$ and $\xi_-$ that are approaching the unit circle and $\xi_+=\xi_-^*$ (orange for $\tau=2.3$ and red for $\tau=2.35$).  Other eigenvalues of $ \hat{\mathfrak{S}}$, in the unit disk, are colored blue, they will not contribute to the long-time limit as they are further away from the unit circle. Similar to Fig. \ref{fig:G3charge} some of the charges in this system are zero, hence while the number of distinct energy levels is 11 (Table \ref{tab:1}) the number of non zero charges, is 7.}
    \label{fig:triple apd}
\end{figure*}

\begin{figure}
    \centering
    \includegraphics[width=1\linewidth]{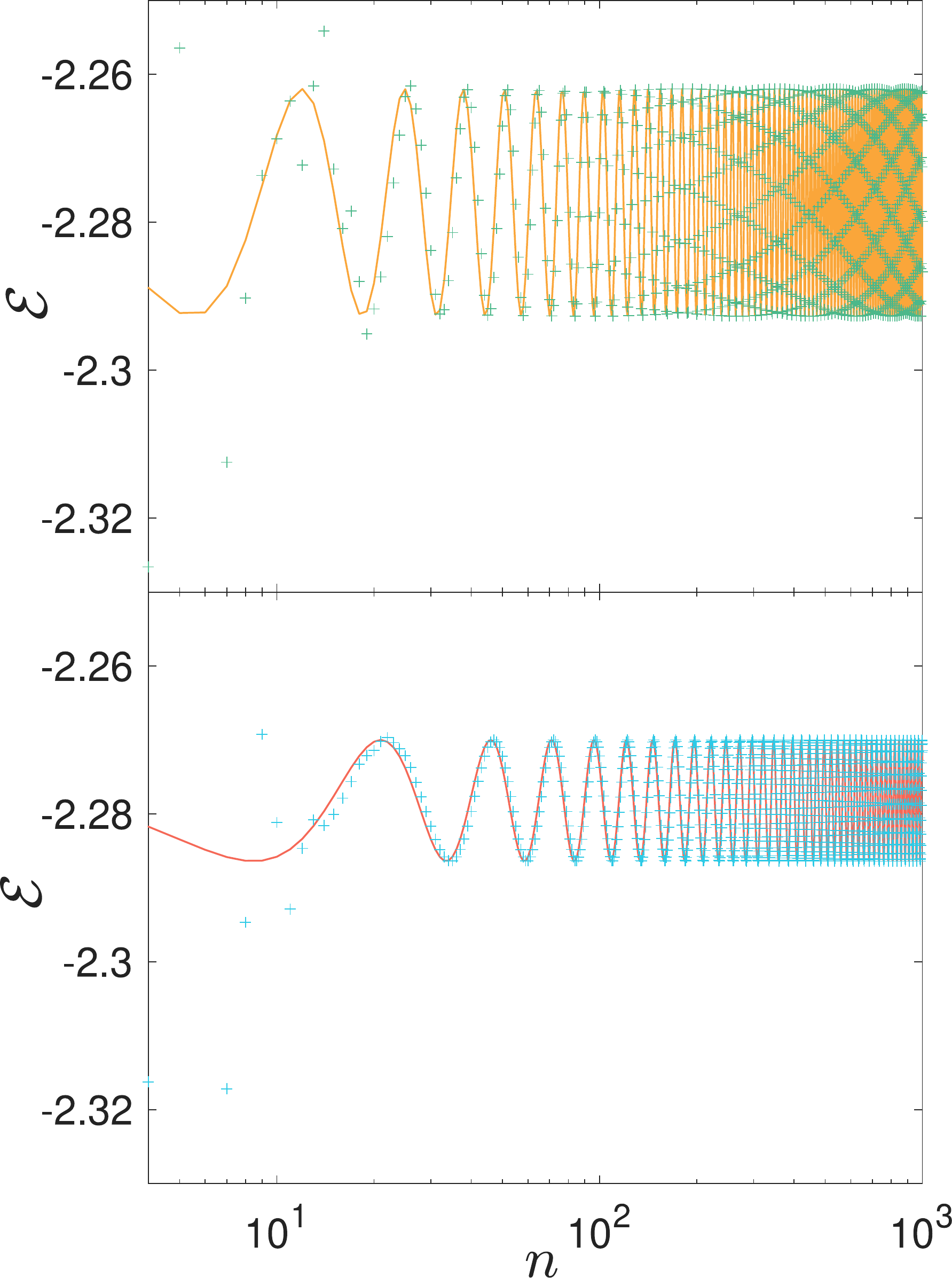}
    \caption{The mean energy of $G_3$ tree system versus the measurement steps $n$. Initially, the system is at its ground state. The measurements drive the mean energy of the system periodically. The upper plot is for $\tau=2.3$, where the green crosses are numerical simulations and the orange line is the theory. The lower plot is for $\tau=2.35$, and the theory (red line) perfectly matches the numerical simulations (blue pentagrams). The oscillation frequency at $\tau=2.3$ is faster than the frequency at $\tau=2.35$, because the relative phase of $\xi_+$ and $\xi_-$ at $\tau=2.3$ is larger than the phase at $\tau=2.35$ (see Fig. \ref{fig:triple apd}). }
    \label{fig:triple}
\end{figure}

An interesting  case is found when three charges merge on the unit circle. In this case we will have two eigenvalues of the survival operator $\hat{\mathfrak{S}}$ in the vicinity of these three charges, and in that sense this example is different from those considered in the previous two subsections. Once we have two eigenvalues which have the same magnitude in absolute value sense, we expect to find  oscillatory behavior, induced by the phase differences, which in turn are controlled in principle by the measurements time $\tau$.  For simplicity we consider systems with commensurate energy levels, the three merging energy levels are $E_0=0$ and $E_{\pm}=\pm E$ with phases $e^0$ and $e^{\pm i E \tau} = e^{ i 2 \pi k \pm i \delta}$, where $k$ is an integer and $\delta$ is our small parameter. This configuration of charges or phases yields two complex eigenvalues denoted $\xi_{\pm}=r_{\pm} e^{i \theta_{\pm}}$. They are located in the vicinity of the unit circle, as expected from basic electrostatics. We denote $p_0=|\langle E_0|\psi_{\rm d}\rangle|^2$ and $p=|\langle E_{\pm}|\psi_{\rm d}\rangle|^2$. As shown in Appendix \ref{apd:triple}, a third-order expansion of Eq. (\ref{charge}) in $\delta$ yields the $\xi_+$ and $\xi_-$ \cite{Yin2019} (see Eqs. (\ref{eq:c1}) and (\ref{eq:c2})).  We find $\xi_+=\xi_-^*$ up to the order $O(\delta^2)$. As mentioned, such a system, will yield quantum dynamics even in the long-time limit, since other eigenvalues of the survival operator are decaying faster, as they are smaller, which is due to the fact that the  three  merging charges are situated one next to each other. In principle the life time of these quantum oscillations depends on the other background charges, that can break symmetry (to higher order in delta) creating a situation where one of the eigenvalues $\xi_+$ or $\xi_-$ is actually closer to the unit circle.

\textbf{Symmetric background:} For symmetric background charges, we have $r_+=r_-$ and $\theta_+=-\theta_-$, and hence, according to our theory, the measurements induce dynamics forever. From Eq. (\ref{eq:quantum dynamics}), the final state is determined by the states $|\xi_+^R\rangle$ and $|\xi_-^R\rangle$. Using Eq. (\ref{righte}), we have:
\begin{equation}
	|\xi_+^R\rangle \sim N \bigg(\frac{\langle E_-|\psi_{\rm d} \rangle}{{\cal C}-1}|E_-\rangle+\frac{\langle E_+|\psi_{\rm d} \rangle}{{\cal C}+1}|E_+\rangle+ \frac{\langle E_0|\psi_{\rm d} \rangle}{{\cal C}}|E_0\rangle \bigg),
\end{equation}
where ${\cal C}=\sqrt{p_0/(p_0+2 p)}$. Similarly, $|\xi_-^R\rangle$ reads:
\begin{equation}
	|\xi_-^R\rangle \sim N \bigg(\frac{\langle E_-|\psi_{\rm d} \rangle}{{\cal C}+1}|E_-\rangle+\frac{\langle E_+|\psi_{\rm d} \rangle}{{\cal C}-1}|E_+\rangle+ \frac{\langle E_0|\psi_{\rm d} \rangle}{{\cal C}}|E_0\rangle \bigg).
\end{equation}
Using Eq. (\ref{eq:quantum dynamics}), since $\theta_+=-\theta_-=\theta$, the final state of the system is $|\psi_f\rangle= N(a_1 e^{i n \theta} |\xi_+^R\rangle+a_2 e^{-i n \theta} |\xi_-^R\rangle)$, where $a_1$ and $a_2$ are overlaps with the initial state. Specially, we choose $a_1=a_2$, which leads to
\begin{equation}
\begin{aligned}
	|\psi_f\rangle \sim N\{&{\cal D}(n)\langle E_-|\psi_{\rm d} \rangle|E_-\rangle+{\cal D}^*(n)\langle E_+|\psi_{\rm d} \rangle|E_+\rangle\\ 
	&-2\cos(n \theta)\langle E_0|\psi_{\rm d} \rangle |E_0\rangle\},
\end{aligned}
\end{equation}
where ${\cal D}(n)=[p_0 \cos(n \theta)+i \sqrt{p_0(p_0+2p)}\sin(n \theta)]/p$ and $\theta$ is defined by $\xi_+=r_+ e^{i \theta_+}=r e^{i \theta}$. Since the first order approximation of $\xi_{\pm}$ is totally imaginary, when $n \ll 1/\delta^2$, we can express $\theta \sim A \delta$, ($A=\sqrt{p_0/(p_0+2p})$, see Eq. (\ref{eq:c3})).

\textbf{Random background:} When the background charges are not strictly symmetric,  $|\xi_+| \neq |\xi_-|$. Then, the quantum dynamics will decay away for very large-$n$ and the system will go to a fixed state determined by the larger of $|\xi_+|$ and $|\xi_-|$. Nevertheless, the system will exhibit nontrivial dynamics in a certain time regime because of the charge configuration we set (the three charges we consider are symmetric).   So the theoretical question is, how long will the quantum dynamics last? From the expressions of $\xi_+$ and $\xi_-$, we see that  $|\xi_+|=|\xi_-|$ up to order $O(\delta^2)$. The background effect comes in order $O(\delta^3)$. So, at least, the quantum dynamics will last until $n \sim 1/\delta^2$. When the number of measurements becomes large than $1/\delta^2$, the system gradually goes to a fixed state due to the symmetry breaking.

\subsubsection{Example}

To demonstrate the triple-charge theory, we use the glued binary $G_3$ tree example and tune the measurement time interval $\tau$.   As shown in Fig. \ref{fig:triple apd}, when $\tau=2.3$ and $\tau=2.35$, we have three charges that are close to each other and far away from other charges (the background charges). As a result, there are two eigenvalues ($\xi_+$ and $\xi_-$) of the survival operator that are near the unit circle (orange for $\tau=2.3$ and red for $\tau=2.35$ in Fig. \ref{fig:triple apd}).  Because of the symmetry of the system, they also have the same absolute value,  so $|\xi_+|=|\xi_-| \simeq 1 $. The exact numerical values, when $\tau=2.3$, are $|\xi_+|=|\xi_-|=0.9873$, while using Eqs. (\ref{eq:c1}) and (\ref{eq:c2}), we have $|\xi_+| \simeq |\xi_-| \simeq 0.9869$.

The triple-charge configuration leads to persistent quantum dynamics. For instance, the mean energy of the system is driven periodically due to the measurements. Here, we choose the initial state to be the ground state, where the effects of the dark states can be neglected. As shown in Fig. \ref{fig:triple},  the mean energy of the $G_3$ system oscillates periodically with the number of measurement steps, $n$, and our theory (lines) fits the numerical simulations (crosses). More importantly, the oscillation frequency of the mean energy changes when we change $\tau$. As shown in the figure, the oscillation frequency at $\tau=2.3$ is faster than the frequency at $\tau=2.35$. This follows from the charge picture, as the relative phase, i.e., $\text{Arg}(\xi_+) - \text{Arg}(\xi_-)$, at $\tau=2.3$ is larger than that at $\tau=2.35$ (see Fig. \ref{fig:triple apd}).  The oscillation frequency is controlled by this relative phase, and hence the frequency at $\tau=2.3$ is faster.

\subsection{Quantum Zeno regime}
 As we increase the number of charges merging to the vicinity  of a point on the unit circle, more eigenvalues of the survival operator approach the unit circle, and they are also all close to each other. An example is the quantum Zeno regime \cite{Misra1977,Fischer2001,Schfer2014,FriedmanE,Lahiri2019,Belan2020,Dubey2020,Thiel2020Zeno}, where $\tau \sim 0$ and all phases $\exp(-iE_k \tau )$ coalesce (as an example, see Fig. \ref{fig:threeabs}(a)). In the quantum Zeno regime, due to the fast measurements, the dynamics of the monitored system is slowed down \cite{Thiel2020Zeno} and our goal is to characterize this behavior. In general, we have a set of eigenvalues of the survival operator (excluding the trivial one on zero) and we consider their absolute value $|\xi_1|, |\xi_2|, \cdots ,|\xi_w|$, which are approaching unity as $\tau \sim 0$. The fastest decay mode is given by the minimum of the set $|\xi_{min}|$. Our goal is to find a rough estimate for this eigenvalue, controlling the rate of the  decay of this component $|\xi_\textit{min}|^n$, see Eq. (\ref{final}).

Basic electrostatics tells us that all the stationary points are located in the convex hull of the charges \cite{Gruenbaum2013}, the area of which vanishes as $\tau \sim 0$. We use this to our advantage and obtain a lower bound for $|\xi|$:
\begin{equation}
	|\xi_{min}| \geq  \cos( \Delta E \tau/2),
	\label{eq:56}
\end{equation}
where $\Delta E$ is the width of the energy spectrum, namely $\Delta E =E_{max}-E_{min}$. An example of Eq. (\ref{eq:56}) is Eq. (\ref{eq:zeno cos}) for the three-level system, which is demonstrated in Fig. \ref{fig:threeabs}(f). Then we let $|\xi_{min}|^{n_b} = e^{-1}$, which leads to $n_b = -1/ \ln [\cos( \Delta E \tau/2)]$. The lower bound of the evolution time $t_b = n_b \tau$ is:
\begin{equation}
	t_b \sim \frac{-\tau}{\ln [\cos( \Delta E \tau/2)]}  \sim \frac{8}{(\Delta E)^2 \tau}.
	\label{eq:47}
\end{equation}
From Eq. (\ref{eq:47}), the evolution of the system is repressed, and the lower bound $t_b$ is proportional to $\tau^{-1}$.  The width of the energy band also affects this lower bound. The bound of the number of measurements then is $n_b = t_b/\tau$, which is proportional to $\tau^{-2}$. So when $\tau \sim 0$, the relaxation is therefore extremely slow, see example in Fig. \ref{fig:three4}(a).

\section{Exceptional points}\label{sec:ep}

Previously in Sec. \ref{sec:two-level} we presented a simple example of an exceptional point of a two-level system. We now briefly explain  how similar effects can be found in larger systems. In particular, we are searching for cases where all the eigenvalues of survival operator  $\xi = 0$. Then, clearly, the right hand side of our main Eq. (\ref{final}) is equal to zero and the whole approach is invalid. As mentioned, this corresponds to a situation where we cannot satisfy the condition of null measurements to begin with.

The basic question is how to construct such systems. Here the charge picture is very useful. Consider a three-level system. Then if we have three charges of the same  magnitude on \ang{-120}, \ang{0}, and \ang{120} on the unit circle, clearly from symmetry all the eigenvalues of $\hat{\mathfrak{S}}$ are zero. This is because the stationary point of this charge configuration is in the centre of the unit disk.  Assume this system has energy levels $-E, 0, E$,  further assume that corresponding wave functions have the same overlaps with the detected state (so the charges are the same). In this case, if we choose $E \tau = 2 \pi/3 + 2 \pi k$ with $k$ an integer, we get a charge configuration that will exhibit the desired result.  It is now rather easy to construct a Hamiltonian that meets this demand, and we present an example below. In fact, this method can be extended to systems beyond  three or two-level systems rather easily. 

We also note that one may have exceptional points that are not zero, i.e. $0 \textless |\xi_i| = |\xi_j| \textless 1$. We do not treat that case here, but leave it for future work.

{\bf Remark:} In the dark subspace, the eigenvalues $\exp(- i E_k \tau)$ of the survival operator $\hat{\mathfrak{S}}$ is $(g_k-1)$-fold degenerate. When $g_k \geq 3$, using Eq. (\ref{darkstateeq}), there are two or more eigenstates $|\delta_{k,i}\rangle$ of the survival operator $\hat{\mathfrak{S}}$. As they are constructed by a Gram Schmidt procedure, these dark states are orthogonal.   Using Eqs. (\ref{righte}) and (\ref{eq:lefteigen}),  for the eigenstates of $\hat{\mathfrak{S}}$ that correspond to the eigenvalues that lie in the unit disk, i.e., $0\textless |\xi| \textless 1$, when two $\xi$s coalesce,  both left and right eigenvectors become parallel. Thus we only have one eigenvector instead of two.   

\subsection{Example}

Here we find the Hamiltonian of the three-level system with the approach mentioned above. As mentioned, we want the three charges to be located on angles \ang{-120}, \ang{0}, and \ang{120}. This can be easily achieved by choosing equally spaced energy levels, for instance, $E_0 =-\gamma$, $E_1 = 0$, and $E_2=\gamma$, where $\gamma \textgreater 0$. When $\tau \gamma = 2\pi/3 + 2\pi k $, we get the desired charge configuration. 

The second step is to have all three charges with equal magnitude, i.e. $p_0=p_1=p_2=1/3$. We choose the detection state $|\psi_{\rm d}\rangle=(1,0,0)^T$. We then construct a group of orthogonal eigenstates of $H$ that $|\langle \psi_{\rm d}|E_i\rangle|^2 =1/3$. A possible choice of such eigenstates $|E_i\rangle$ is  $|E_0\rangle = (1/\sqrt{3},-1/\sqrt{6},1/\sqrt{2} )^T$, $|E_1 \rangle = (-1/\sqrt{3},1/\sqrt{6},1/\sqrt{2})^T$, and $|E_2 \rangle = (1/\sqrt{3},\sqrt{2/3},0)^T$. Such eigenstates of $H$ have identical overlaps with the detection state, and hence all the charges have the same magnitude.

The last step is to find the concrete form of $H$. Since $|E_i\rangle$ is an eigenstate of $H$ with eigenvalues $E_i$, we have $H|E_i\rangle = E_i|E_i\rangle$, which leads to
\begin{equation} H=-\gamma \left(
\begin{matrix}
  0 & -1/\sqrt{2} & 1/\sqrt{6} \\
  -1/\sqrt{2} & -1/2 & -1/(2 \sqrt{3}) \\
  1/\sqrt{6} & -1/(2\sqrt{3}) & 1/2
\end{matrix}\right).
\end{equation}
For a quantum system with such a Hamiltonian, when $\tau \gamma = 2\pi/3 + 2\pi k $, we cannot satisfy the condition of null measurements. In other words, the system is detected with probability one by the local measurements. $\xi=0$ is an exceptional point of the survival operator of order three. With the same procedure, this can be generalized for larger systems.

\section{Summary}\label{sec:summary}

We have investigated the properties of the quantum survival operator $\hat{\mathfrak{S}}$, which in turn gives the state of the system after a large number of conditional measurements. We have classified five types of generic behaviors:
\begin{itemize}
\item[1.] In the presence of symmetry and hence a spectrum which is degenerate, the system will exhibit dynamics determined by the energy levels of the systems. However, only the degenerate energy levels participate. Of course, this is the case under the condition that initially $|\psi_{{\rm in}}\rangle $  has some overlap with the dark part of the Hilbert space. A non-degenerate energy level may contribute only if it is orthogonal to the detected state. 
\item[2.] In the absence of a dark subspace, for example, for systems with no-degeneracy, e.g. disordered, interacting or chaotic systems, typically the final state of the system is unique. This means that in the long-time limit, we have no dynamics at all, and we can say that the system reaches a kind of equilibrium, induced by the repeated measurement. This state corresponds the eigenvalue of the survival operator closest to the unit circle. In the second part of the paper we developed tools to identify this state.  
\item[3.] Under certain conditions, we find a pair of eigenvalues of $\hat{\mathfrak{S}}$ which together are also the largest in magnitude. We find that these eigenvalues then have a relative phase that is controlled by the rate $1/\tau$. In this case, the system exhibits oscillation which are controlled by the measurements, and not by the energy levels alone (unlike standard quantum mechanics).
\item[4.] In the Zeno limit, all the eigenvalues of $\hat{\mathfrak{S}}$ approach the unit circle. This means that oscillations are effectively undetectable, and the relaxations are slow (as well known). We have presented a lower bound for the eigenvalues, showing their vicinity to the unit circle.
\item[5.] An interesting case is a situation where all the eigenvalues of $\hat{\mathfrak{S}}$ are equal to zero. This corresponds to widely investigated exceptional points of non-Hermitian systems. The physics in this case implies that the condition of null measurements is not realizable.  Rather the system is detected with probability one, and hence the condition we impose is violated. The classical charge picture was proven to be very useful here, in the sense that we can exploit the symmetry of the charge picture to direct the eigenvalues of $\hat{\mathfrak{S}}$ to a unique stable point in the centre of the unit disk, and hence exhibit the exceptional physics.
\end{itemize}

We have considered the examples of two, and three-level systems, as well as a binary glued tree. We use the mean energy of the system to characterize the state of the systems, while other observables are also possible.  The two-level system was used to exhibit an exceptional point  (effect 5). However, with a two-level system one cannot detect the measurement induced dynamics (effect 3) simply because in a two-level system one has only one non-trivial eigenvalue of the survival operator. In a three-level $V$-shaped  artificial atom (see Fig. \ref{fig:three}), we detected shelving and effect 2 (see Fig. \ref{fig:three-energy}), which is actually well known \cite{Plenio1998}. The theory developed here, in particular the charge theory, makes it possible to realize the long-time behavior using an intuitive electrostatic analogy. For example, a single weak charge, corresponding to a weak overlap of a stationary energy  with the detected state, or merging charges (phases) on the unit circle, imply an eigenvalue of $\hat{\mathfrak{S}}$  which is close to unity in magnitude, and its precise value can be estimated in generality. Maybe more interesting is the merging of three charges on the unit circle, since here we get two eigenvalues of $\hat{\mathfrak{S}}$  approaching the unit charges, and then we get the dynamical effect due to the phase difference (effect 3). These effects was explored with the glued tree example (see Fig. \ref{fig:triple}). Clearly the high degree of symmetry in this case, and the energy degeneracy of the system, imply a large dark subspace (unlike the three-level systems under study) and hence this type of example exhibits physical behaviors drastically different from the simple three-level systems. As mentioned, the charge picture was used also to find a lower bound for the Zeno limit, and helps tremendously in the identification of exceptional points.

\section{Acknowledgements}
The support of Israel Science Foundation's grant 1898/17 as well as the support by the Julian Schwinger Foundation (KZ) are acknowledged.

\appendix

\section{Matrix determinant  lemma}\label{ap:A}

In this section, we present the details for the calculation using the matrix determinant lemma. The formula of the matrix determinant lemma is:
\begin{equation}
	\text{det}(A+uv^T)=(1+v^T A^{-1} u)\text{det}(A),
\end{equation}
where $A$ is an invertible square matrix and $u$, $v$ are column vectors. For the eigenvalues of the survival operator $ \hat{\mathfrak{S}}$, we have:
\begin{equation}
	\text{det}(\xi- \hat{\mathfrak{S}})=\text{det}[\xi-\hat{U}(\tau)+|\psi_{\rm d}\rangle \langle \psi_{\rm d}| \hat{U}(\tau)] .
\end{equation}
We let $A=\xi-\hat{U}(\tau)$, $u= |\psi_{\rm d}\rangle$, and $v^T= \langle \psi_{\rm d}| \hat{U}(\tau)$, which leads to:
\begin{equation}
	\text{det}(\xi- \hat{\mathfrak{S}})=\text{det}[\xi-\hat{U}(\tau)]\langle \psi_{\rm d}|[\xi-\hat{U}(\tau)]^{-1}|\psi_{\rm d}\rangle \xi.
\end{equation}
This is the formula we used in the main text.

\section{Calculation for the right eigenvectors of survival operator with eigenvalues in the unit disk}\label{calculate the xi}
In this section, we present the details for Eq. (\ref{righte}) in the main text. As discussed in the main text, we expand the right eigenstates in the bright subspace.
\begin{equation}
	|\xi^R\rangle=\sum_{\{B\}}a_{i} |\beta_i\rangle,
	\label{ap21}
\end{equation}
where $a_i$ is the index we are looking for and $\{B\}$ represents the summation in the bright subspace. Substituting Eq. (\ref{ap21}) into Eq. (\ref{eq:eigenG}), we have:
\begin{align}
	\hat{\mathfrak{S}}|\xi^R\rangle &=\sum_{\{B\}}a_{i} (1-D) \hat{U}(\tau)  |\beta_i\rangle, \\
	&= \sum_{\{B\}}a_{i} (1-D) e^{- i \tau E_i} |\beta_i\rangle, \label{15}\\ 
    &= \xi \sum_{\{B\}}a_{i} |\beta_i\rangle. \label{16}
\end{align}
Multiplying Eqs. (\ref{15}, \ref{16}) $\langle \beta_j|$, we have:
\begin{equation}
	\xi a_{j}= \sum_{\{B\}}a_{i} e^{- i \tau E_i} \langle \beta_j|(1-D)|\beta_i\rangle.
	\label{ap25}
\end{equation}
Following the definition of the charges, the matrix elements on the rhs of Eq. (\ref{ap25}) can be simplified.
\begin{equation}
	\langle \beta_j|(1-D)|\beta_i\rangle= \left\{
	\begin{aligned}
		 1-\langle \beta_j|D|\beta_j\rangle=1- p_j \qquad &i=j\\
		- \langle \beta_j|D|\beta_i\rangle = -\sqrt{p_j p_i}  \qquad &i\neq j
	\end{aligned}
	\right ..
	\label{ap26} 
\end{equation}
Substituting Eq. (\ref{ap26}) into Eq. (\ref{ap25}), we have
\begin{equation}
    \sum_{\{B\},i\neq j} a_{i} e^{- i \tau E_i}\sqrt{p_i}=a_{j}\frac{[ e^{-i \tau E_j} (1-p_j)-\xi_l]}{\sqrt{p_j}} .
    \label{aije}
\end{equation}
Now we define $ b_{i}=a_{i} e^{- i \tau E_i}\sqrt{p_i}$, so $a_{i}=b_{i}/(e^{- i \tau E_i}\sqrt{p_i})$. We also define $\zeta_{j}=1-(1-e^{i \tau E_j}\xi)/ p_j$. Eq. (\ref{aije}) then can be simplified as:
\begin{equation}
    \sum_{\{B\},i\neq j} b_{i}=b_{j}\zeta_j.
    \label{ap28}
\end{equation}
For $j=1, 2, \cdots, w$, Eq. (\ref{ap28}) is an equation set, which contains $w$ terms. We want to rewrite this equation set into the matrix form. We define the vector $B$ as:
\begin{equation}
	B^{\dagger}= (b_{1}^*, b_{2}^*, \cdots, b_{w}^*).
\end{equation}
Then the matrix is:
\begin{equation}M=\left \{
\begin{matrix}
  \zeta_{1} & 1   &  1 & \cdots &  1 \\
  1 & \zeta_{2} &   1 & \cdots &  1  \\
  1 &  1 & \zeta_{3} & \cdots &   1\\
 \vdots & \vdots & \vdots & \vdots & \vdots \\
  1 & 1 &  1  & \cdots &  \zeta_{w}.
\end{matrix}
\right \}.
\label{ap2m}
\end{equation}
Eq. (\ref{ap28}) then can be rewritten as:
\begin{equation}
	M B=0.
	\label{ap211}
\end{equation}
Using Eq. (\ref{ap211}) or Eq. (\ref{ap28}), we have:
\begin{equation}
    b_{i}=b_{1} \frac{\zeta_{1}-1}{\zeta_{i}-1}.
    \label{ap12Feb}
\end{equation}
Substituting Eq. (\ref{ap12Feb}) into the expressions for $b_i$ and $\zeta_i$, we get the expression for $a_i$:
\begin{equation}
    a_{i}=a_{1}\sqrt{\frac{p_i}{p_1}}\frac{e^{-i \tau E_1}-\xi}{e^{-i \tau E_i}-\xi}=
    \frac{a_{1}(e^{-i \tau E_1}-\xi)}{\sqrt{p_1}} \frac{\sqrt{p_i}}{e^{-i \tau E_i}-\xi}.
    \label{ap213}
\end{equation}
From Eq. (\ref{ap213}), we see that $a_2, a_3, \cdots, a_w$ can be expressed by the $a_1$. We can choose $a_1$ freely, and in the end it is a global constant that can be neglected by the normalization. Here we choose 
\begin{equation}
	a_1= \frac{\sqrt{p_1}}{e^{-i \tau E_1 }-\xi},\ \text{then}\ a_i= \frac{\sqrt{p_i}}{e^{-i \tau E_i}-\xi}.
\end{equation}
Using Eq. (\ref{ap21}), we have:
\begin{equation}
	 |\xi^R\rangle =  \sum_{\{B\}} \frac{\sqrt{p_i}}{e^{-i \tau E_i}-\xi} |\beta_i\rangle.
	 \label{ap215}
\end{equation}
Using Eq. (\ref{bright}), the bright states $|\beta_i\rangle= \hat{P}_i|\psi_{\rm d}\rangle/\sqrt{p_i}$. So together with Eq. (\ref{ap215}) we have:
\begin{equation}
     |\xi^R\rangle =  \sum_{i=1}^w \frac{\hat{P_i}}{e^{-i \tau E_i}-\xi} |\psi_{\rm d}\rangle.
\end{equation}

\section{Relation between right and left eigenstates}\label{apd:relation}
In this section we derive a relation that connects the right eigenvectors to the left eigenvectors. Using Eqs. (\ref{eq:rightG}) and (\ref{eq:leftG}), we find the right and left eigenstates can be related. 
\begin{equation}
	\hat{U}(\tau)|\xi_{\iota}^R\rangle = |\xi_{\iota}^L\rangle^* ,
	\label{apeq:20}
\end{equation}
here $*$ is the complex conjugate. For a two-level system, if $|\xi_{\iota}^L\rangle=\{a+i b, c + i d\}^T$ ($a$, $b$, $c$, $d$ are real numbers), then $|\xi_{\iota}^L\rangle^*=\{a-i b, c - i d\}^T$.     Eq. (\ref{apeq:20}) shows that the evolution operator $\hat{U}(\tau)$ maps the right eigenstate of survival operator $\hat{\mathfrak{S}}$ to the corresponding left one.

\section{Triple-charge theory}\label{apd:triple}

In this section we present the approximate expressions of $\xi_+$ and $\xi_-$ for the triple-charge theory. The perturbation approach applied here is that we perform a third-order expansion of Eq. (\ref{charge}) in small parameter $\delta$. The eigenvalues $\xi_+$ and $\xi_-$ then reads:
\begin{equation}
	\xi_+=r_+ e^{i \theta_+} \sim 1 + i A \delta-B\delta^2+O(\delta^3);
	\label{eq:c1}
\end{equation}
\begin{equation}
	\xi_- =r_- e^{-i \theta_-} \sim 1 - i A \delta-B\delta^2+O(\delta^3),
	\label{eq:c2}
\end{equation}
where
\begin{equation}
	A=\sqrt{\frac{p_0}{p_0+2p} };
	\label{eq:c3}
\end{equation}
\begin{equation}
	B=\frac{p_0+p}{2p_0 + 4 p} -\frac{p}{(p_0+2 p)^2} \sum_{j\neq 0, \pm} \frac{p_j}{1-e^{- i E_j \tau}}.
\end{equation}


%

\end{document}